\documentclass[12pt]{article}
\usepackage{epsfig,amssymb}
\usepackage{latexsym}

\newcommand{\bq}{\begin{equation}}
\newcommand{\eq}{\end{equation}}
\newcommand{\bqa}{\begin{eqnarray}}
\newcommand{\eqa}{\end{eqnarray}}
\newcommand{\ben}{\begin{enumerate}}
\newcommand{\een}{\end{enumerate}}
\newcommand{\bc}{\begin{center}}
\newcommand{\ec}{\end{center}}
\newcommand{\bqb}{\begin{eqnarray*}}
\newcommand{\eqb}{\end{eqnarray*}}

\def\pr#1#2#3{Phys. Rev. ${\bf{#1}}$, #2 (#3)}
\def\prl#1#2#3{Phys. Rev. Lett. ${\bf{#1}}$, #2 (#3)}
\def\pl#1#2#3{Phys. Lett. ${\bf{#1}}$, #2 (#3)}
\def\prep#1#2#3{Phys. Rept. ${\bf{#1}}$, #2 (#3)}

\def\ijmp#1#2#3{Int. J. Mod. Phys. ${\bf{#1}}$, #2 (#3)}

\begin{document}
\pagenumbering{arabic}
\thispagestyle{empty}
\def\thefootnote{\fnsymbol{footnote}}
\setcounter{footnote}{1}

\begin{flushright}
Sept 7,  2015. \\
Corrected version.\\
arXiv: 1507.08827\\
 \end{flushright}

\vspace{2cm}

\begin{center}
{\Large {\bf The Higgs boson structure functions as signposts\\
of new physics}}.\\
 \vspace{1cm}
{\large G.J. Gounaris$^a$ and F.M. Renard$^b$}\\
\vspace{0.2cm}
$^a$Department of Theoretical Physics, Aristotle
University of Thessaloniki,\\
Gr-54124, Thessaloniki, Greece.\\
\vspace{0.2cm}
$^b$Laboratoire Univers et Particules de Montpellier,
UMR 5299\\
Universit\'{e} Montpellier II, Place Eug\`{e}ne Bataillon CC072\\
 F-34095 Montpellier Cedex 5.\\
\end{center}

\vspace*{1.cm}
\begin{center}
{\bf Abstract}
\end{center}

We show that the Higgs boson structure functions observable in the inclusive
process $e^-e^+ \to H + {\rm ``anything"}$, may reveal the presence of anomalous
contributions corresponding to several types of new physics partners,
Higgs boson compositeness, or invisible (dark) matter.
This could be done without making a difficult or even an impossible
experimental analysis of what ``anything" contains.
We give illustrations showing how the shapes of the various structure
functions induced by such  new contributions may differ
from the standard predictions, thus possibly allowing their identification.\\

\vspace{0.5cm}

\def\thefootnote{\arabic{footnote}}
\setcounter{footnote}{0}
\clearpage

\section{Introduction}

The discovery \cite{Higgsdiscov} of the Higgs boson \cite{Higgs}
is a great step towards  the verification of the standard model (SM) \cite{Higgssearch}.
Nevertheless there are several experimental (muon g-2, neutrino masses, dark matter)
and theoretical (hierarchy problem,...) points which are not covered by
 SM \cite{BSMexp}.
Many ways, of very different nature, have been proposed for making
a fruitful extension of SM, like supersymmetry, additional strong sectors, compositeness,
and dark sectors, globally called ``beyond the standard model'' (BSM)  \cite{BSMth}.

Here we want to turn in particular to the  difficulty  in analyzing  the mechanisms
responsible for producing multiparticle or invisible states, in particular
dark matter, in association with the Higgs particle. Indeed,
the Higgs boson may couple to several  new particles  and
may be a portal to such new sectors \cite{Portal}.
So we thought that a possible hint  may arise if we look at the ``Higgs structure functions"
in the \underline{inclusive} process $e^-e^+ \to H(p_H^\mu) + X $, where $p_H^\mu$ denotes the
$H$-four-momentum, $s$ describes  the total  $e^-e^+$ c.m. energy-squared,
 and $X$ stands for   an additional final state  with
two or more particles. We ignore contributions where $X$ is a single particle associated
to  $H$ (for example $e^-e^+\to HZ$); since it is  localized at a single point
near the end of the spectrum, with the c.m. $H$-energy being $p^0_H=(s+m^2_H-m^2_Z)/(2\sqrt{s})$.
Instead,  we are rather interested by the shapes of the structure functions versus $s$ and $p_H$.

At LHC, there may also be other similar  inclusive processes,
where the Higgs-particle is produced in association with new particles.
But in this paper we only consider the $e^-e^+$ process, which
is probably the simplest and  clearest one phenomenologically.

In particular, if the $X$-state mentioned above is generated through an  $s$-channel intermediate state
$(V=\gamma,~Z)$, as in
\bq
e^-e^+ \to V(\gamma ,Z)\to H(p_H^\mu) + X ~~,  \label{BSM-X}
\eq
which is true for the BSM models used here and partly also for SM,
then the general form of the cross section for any $e^-e^+$-polarizations,
 may be described by just three   structure functions $W_i$ $(i=1,2,3)$.
 These $W_i$ functions may then be chosen
 to depend only on $s$ and  the c.m. magnitude of the  $H$ momentum \cite{Wi}.

In Sect.2 we give the  definitions of these functions  and  recall the relations among the three $W_i$,
induced by  the quantum numbers of the particles associated to the $H$ production.
In Sect.3 we show how they can be simply measured through the angular
distribution of the $H$-inclusive cross section in the $e^-e^+$ unpolarized
and polarized cases.
In Sect.4 we compute  their standard contributions, where the 2-body $X$ (``anything") state
associated to  $H$, is due to the usual fermion pairs ($t\bar t$ and the light $f\bar f$
ones) and the $ZH$ bosons.

We treat separately the case $X=W^-W^+$, which can only partly be described
by the form (\ref{BSM-X}); since now  one has to add the t-channel neutrino exchange
amplitude, followed by a $W^{\mp}W^{\pm}H$ coupling.
We also include separately the similar t- and u- channel electron exchange
processes $e^-e^+\to ZZ$ and $e^-e^+\to Z\gamma$  followed by a $ZZH$ coupling.

For simplicity, we  ignore the contributions of the $ZZ$  and $W^-W^+$ fusion
processes  $e^-e^+\to e^-e^+H$ and $e^-e^+\to \nu_e \bar\nu_e H$ (see \cite{ILC}), since
these processes can be considered as a ``background" which
may be subtracted from the data\footnote{See Section 4.},
 before making the analysis  we propose.

In Sect.5 we then give typical examples of possible new physics contributions; i.e.
production of new bosons or fermions emitting the Higgs boson, either in the form of
additional associated two-body contributions of type (\ref{BSM-X}),
or in a form similar to the hadronic parton model in the case of Higgs boson compositeness.
We then show how the structure functions $W_i$ depend on the nature of the associated particles,
and on their parameters (masses, couplings). In particular we show how  the shapes of
the $s$ and $p_H$ dependencies of the various $W_i$ are affected by   the nature of the
associated particles, which are generally difficult to detect directly.
Several illustrations for the various combinations of these structure
functions are given.

Summarizing, the contents of the paper are the following. In Sect.2 we give the precise definition
of the structure functions. In Sect.3 we express their measurability through the angular dependencies
of the polarized and unpolarized cross sections. Sect.4 is devoted to the theoretical computation
and illustrations of the
SM contributions to the structure functions, while Sect.5 concentrates in some
 examples of new physics effects.
Conclusions mentioning possible future developments are given in Sect.6.\\

\section{Structure functions}

The cross section of the process (\ref{BSM-X}) may be written   as \cite{Wi}
\bq
\sigma = {(2\pi)^4\over 2s} \int d\rho L^{\alpha\beta}_{\mu\nu}
H^{\mu\nu}_{\alpha\beta}~~, \label{sigma}
\eq
where $L^{\alpha\beta}_{\mu\nu}$ arise from the square of the  initial
vertex and the $\gamma,Z$ propagators, while
$H^{\mu\nu}_{\alpha\beta}$ comes from the  final part
$\gamma,Z \to H + X$, containing the Higgs structure functions;
$\alpha,\beta$ refer to $\gamma$ or $Z$. Denoting by $p_H^\mu$
the four-momentum of the final $H$-particle, and by  $q^\nu$
the $e^-e^+$-total four-momentum, we define
\bq
d\rho={d_3p_H\over2(2\pi)^3p_H^0}d\rho_X ~~, \label{drho}
\eq
 with $d\rho_X$ containing the phase-space of the final  $X$-state.
  Using this, one writes
\bqa
&& \int d\rho_X H^{\mu\nu}_{\alpha\beta}={1\over(2\pi)^3}W^{\mu\nu}_{\alpha\beta}
=\sum_i I^{\mu\nu}_iW^{\alpha\beta}_i\nonumber\\
&&
=-(g^{\mu\nu}-{q^{\mu}q^{\nu}\over s})W^{\alpha\beta}_1
+(p_H^{\mu}-{p_H.q\over s}q^{\mu})(p_H^{\nu}-{p_H.q\over s}q^{\nu}){W^{\alpha\beta}_2\over m^2_H}
-i\epsilon^{\mu\nu\rho\sigma}p_{H\rho}q_{\sigma}{W^{\alpha\beta}_3\over m^2_H}~,\label{Walphabeta-i-1}
\eqa
where
\bq
I_1=-(g^{\mu\nu}-{q^{\mu}q^{\nu}\over s})~~,~~
I_2={1\over m^2_H}(p_H^{\mu}-{p_H.q\over s}q^{\mu})(p_H^{\nu}-{p_H.q\over s}q^{\nu})~~,~~
I_3={-i\over m^2_H}\epsilon^{\mu\nu\rho\sigma}p_{H\rho}q_{\sigma}~~, \label{I123-forms}
\eq
are the only terms that remain   non vanishing after contracting
with $L^{\alpha\beta}_{\mu\nu}$; compare (\ref{sigma}).
 Note that the H structure functions $W^{\alpha\beta}_i$ in (\ref{Walphabeta-i-1}),
are taken as functions of  $s=q^2$ and the  c.m. magnitude of the  $H$ momentum
\bq
p_H=\frac{\sqrt{[s-(m_H+M_X)^2][s-(m_H-M_X)^2]}}{2\sqrt{s}} ~~, \label{pH}
\eq
 where  $M_X$ denotes the invariant mass of the $X$-state. As a first step we only  consider
two-body $X$-states. At very high energies, multibody $X$ contributions would probably also be needed.\\

\noindent
{\bf $H$+2-body contributions}\\
We start from the transition
$\gamma,Z \to x+x'$ followed by  $x\to x^{\prime\prime} +H$ or $x' \to x^{\prime\prime}+H$, where
 $x,x',x^{\prime\prime}$ can be scalars, fermions or gauge bosons.
For  each $X$,  one squares the sum over all diagram, obtaining  an expression
for $H^{\mu\nu}_{\alpha\beta}$.
We  then compute its corresponding three numerical quantities,
\bq
K^{\alpha\beta}_i=\int d\rho_X H^{\mu\nu}_{\alpha\beta}~I^{\mu\nu}_i~~, \label{Kalphabeta-i}
\eq
by integrating the $I^{\mu\nu}_i$-expressions in (\ref{I123-forms}). Denoting the respective momenta
of $( x^\prime~,~x^{\prime\prime})$ as  $(p'~,~p^{\prime\prime})$, we have
\bq
d\rho_X={1\over4(2\pi)^6}{d_3p'd_3p^{\prime\prime}\over p^{'0}p^{\prime\prime 0}}
\delta_4(q-p_H-p'-p^{\prime\prime}) ~~, \label{drhoX-2}
\eq
which through (\ref{Walphabeta-i-1}) lead to
\bqa
W^{\alpha\beta}_1 &= & {sm^2_H\over2[sm^2_H-(p_H.q)^2]}
\left [{K^{\alpha\beta}_1\over sm^2_H}[sm^2_H-(p_H.q)^2]+K^{\alpha\beta}_2 \right ] ~~, \nonumber \\
W^{\alpha\beta}_2 &= & {s^2m^4_H\over
 2[sm^2_H-(p_H.q)^2]^2}\left [{K^{\alpha\beta}_1\over sm^2_H}[sm^2_H-(p_H.q)^2]
+3K^{\alpha\beta}_2 \right ]  ~~, \nonumber \\
W^{\alpha\beta}_3 &= & {m^4_H\over2[{sm^2_H-(p_H.q)^2}]}K^{\alpha\beta}_3 ~~. \label{Walphabeta-i-2}
\eqa\\

\section{The $e^-e^+$ cross section}

Using (\ref{sigma}-\ref{I123-forms}) and $\alpha=e^2/(4\pi)$, the general inclusive cross section
for polarized $e^{\mp}$ beams  is written as
\bqa
{p_H^0d\sigma (e^-e^+\to H+X) \over  d_3p_H}&=& {\alpha^2\over s^2} \Big \{(1-P_LP'_L)U_1+(P_L-P'_L)U_2
\nonumber \\
&& +P_TP'_T(U_3\cos2\phi+U_4\sin2\phi) \Big \}~~, \label{sigma1}
\eqa
\noindent
where  $P_L$, $P_{L'}$, $P_T$, $P_{T'}$ are the longitudinal and transverse
$e^{\mp}$ beam degrees of polarization  and
\bqa
&&U_1=\left (W^{\gamma\gamma}_1+{p^2_H\over2m^2_H}W^{\gamma\gamma}_2\sin^2\theta \right )
+{s^2(|g^{L}_{Zee}|^2+|g^{R}_{Zee}|^2)\over2(s-m^2_Z)^2}
\left (W^{ZZ}_1+{p^2_H\over2m^2_H}W^{ZZ}_2\sin^2\theta \right )
\nonumber\\
&&-Re\Bigg \{{s(g^{L*}_{Zee}+g^{R*}_{Zee})\over (s-m^2_Z)}\Big (W^{\gamma Z}_1+{p^2_H\over2m^2_H}
W^{\gamma Z}_2\sin^2\theta \Big )\Bigg \}
\nonumber\\
&&+{s^2(|g^{L}_{Zee}|^2-|g^{R}_{Zee}|^2)p_H\sqrt{s}\over 2(s-m^2_Z)^2m^2_H}
W^{ZZ}_3\cos\theta
+Re\Bigg\{{s(g^{R*}_{Zee}-g^{L*}_{Zee})p_H\sqrt{s}\over (s-m^2_Z)m^2_H}W^{\gamma Z}_3\Bigg \}\cos\theta
~, \nonumber \\
&&U_2={s^2(|g^{L}_{Zee}|^2-|g^{R}_{Zee}|^2)\over2(s-m^2_Z)^2}
\left (W^{ZZ}_1+{p^2_H\over2m^2_H}W^{ZZ}_2\sin^2\theta \right )
\nonumber \\
&&+Re\left \{{s(g^{R*}_{Zee}-g^{L*}_{Zee})\over (s-m^2_Z)}\left (W^{\gamma Z}_1+{p^2_H\over2m^2_H}
W^{\gamma Z}_2 \sin^2\theta \right )\right \}
\nonumber\\
&&+{s^2(|g^{L}_{Zee}|^2+|g^{R}_{Zee}|^2)p_H\sqrt{s}\over 2(s-m^2_Z)^2m^2_H}W^{ZZ}_3\cos\theta
-Re\left \{{s(g^{L*}_{Zee}+g^{R*}_{Zee})p_H\sqrt{s}\over (s-m^2_Z)m^2_H}W^{\gamma Z}_3\right \}
\cos\theta ~, \nonumber \\
&&U_3={-p^2_H\over2m^2_H}W^{\gamma\gamma}_2\sin^2\theta
-2{ s^2Re\{g^{L}_{Zee}g^{R}_{Zee}\}p^2_H\over2m^2_H(s-m^2_Z)^2}W^{ZZ}_2\sin^2\theta
\nonumber\\
&&+{p^2_HRe\{s(g^{L*}_{Zee}+g^{R*}_{Zee})W^{\gamma Z}_2\}\over 2m^2_H(s-m^2_Z)}
\sin^2\theta ~, \nonumber \\
&& U_4=-{p^2_HIm\{s(g^{L*}_{Zee}-g^{R*}_{Zee})W^{\gamma Z}_2\}\over 2m^2_H(s-m^2_Z)}
\sin^2\theta ~~. \label{U1234-forms}
\eqa
Note that $W^{\gamma \gamma}_i$, $W^{ZZ}_i$ are real, while
$W^{\gamma Z}_i=W^{Z\gamma *}_i$ can be complex.
Consequently, at tree level and for  real photon and $Z$
couplings, we have  $U_4=0$.   Non vanishing contributions to $U_4$ may only come
from loop corrections or  effective $Zee$-form factors.

In Sections 4 and 5 we discuss   illustrations  of SM and new physics
contributions to the $W_i$ structure functions
arising  in the unpolarized case, as well as in the cases
of $e^\mp$ beams with only longitudinal or only transverse polarizations;
for simplicity we restrict to real couplings. We emphasize that such illustrations  apply
only to $X$-states appearing in transitions of the type (\ref{BSM-X}).

Therefore, for the formalism based on (\ref{U1234-forms}) to apply,
 special care   must be taken, so that the experimental measurements assure
the exclusion of $X$-states not-generated  by $\gamma, ~Z \to H+X$, but rather through
$t$ or $u$ channel exchanges in $e^-e^+\to H+X$. In the present work such $X$-states
only appear in the SM contribution, and they are  calculated directly; see Section 4. \\

\noindent
{\bf $U_1$ determination through unpolarized beams}.\\
Using (\ref{sigma1}), the differential cross section for unpolarized beams may
be written as
\bq
{d\sigma(e^-e^+\to H+X)\over dp_Hd\cos\theta_H}={2\pi p_H^2\over p^0_H}
{p_H^0d\sigma (e^-e^+\to H+X) \over  d_3p_H}  ~~, \label{sigma2}
\eq
where only $U_1$  of (\ref{U1234-forms}) contributes. The explicit expression thus obtained
consists of  a constant (angular independent) term
\bq
V_1=W^{\gamma\gamma}_1
+{s^2(g^{L2}_{Zee}+g^{R2}_{Zee})\over2(s-m^2_Z)^2}W^{ZZ}_1
-{s(g^{L}_{Zee}+g^{R}_{Zee})\over (s-m^2_Z)}W^{\gamma Z}_1 ~~, \label{V1-form}
\eq
a term  proportional  to $\sin^2\theta$
\bq
V_2={p^2_H\over2m^2_H}\sin^2\theta \left [W^{\gamma\gamma}_2
+{s^2(g^{L2}_{Zee}+g^{R2}_{Zee})\over2(s-m^2_Z)^2}W^{ZZ}_2
-{s(g^{L}_{Zee}+g^{R}_{Zee})\over (s-m^2_Z)}
W^{\gamma Z}_2 \right ]  ~~, \label{V2-form}
\eq
and a term proportional  to $\cos\theta$
\bq
V_3=\left [{s^2(g^{L2}_{Zee}-g^{R2}_{Zee})p_H\sqrt{s}\over 2(s-m^2_Z)^2m^2_H}W^{ZZ}_3+{s(g^{R}_{Zee}
-g^{L}_{Zee})p_H\sqrt{s}\over (s-m^2_Z)m^2_H}W^{\gamma Z}_3 \right ]\cos\theta  ~~. \label{V3-form}
\eq\\

\noindent
{\bf Longitudinally polarized beams may be used to determine $U_2$}.\\
Using (\ref{sigma1}, \ref{U1234-forms}), this is found to consist of  the angle-independent term
\bq
V_4={s^2(g^{L2}_{Zee}-g^{R2}_{Zee})\over2(s-m^2_Z)^2}W^{ZZ}_1
+{s(g^{R}_{Zee}-g^{L}_{Zee})\over (s-m^2_Z)}W^{\gamma Z}_1 ~~, \label{V4-form}
\eq
a term proportional  to $\sin^2\theta$
\bq
V_5={p^2_H\over 2m^2_H}\sin^2\theta \left [{s^2(g^{L2}_{Zee}-g^{R2}_{Zee})\over2(s-m^2_Z)^2}
W^{ZZ}_2+{s(g^{R}_{Zee}-g^{L}_{Zee})\over (s-m^2_Z)}
W^{\gamma Z}_2 \right ] ~~, \label{V5-form}
\eq
and a term proportional  to $\cos\theta$
\bq
V_6=\cos\theta{p_H\sqrt{s}\over m^2_H}\left [{s^2(g^{L2}_{Zee}+g^{R2}_{Zee})\over 2(s-m^2_Z)^2}W^{ZZ}_3
-{s(g^{L}_{Zee}+g^{R}_{Zee})\over (s-m^2_Z)}W^{\gamma Z}_3 \right ] ~~. \label{V6-form}
\eq\\

\noindent
{\bf Transversally polarized beams may be used to determine $U_3,~U_4$}:\\
Using again  (\ref{sigma1}, \ref{U1234-forms}) and real couplings so that $U_4=0$,
the obtained cross section is fully determined by
\bqa
V_7 =U_3&=&p^2_H  \sin^2\theta \Bigg [{-W^{\gamma\gamma}_2 \over2m^2_H}
-2{ s^2Re\{g^{L}_{Zee}g^{R}_{Zee}\}\over2m^2_H(s-m^2_Z)^2}W^{ZZ}_2
\nonumber \\
&& +{Re\{s(g^{L*}_{Zee}+g^{R*}_{Zee})W^{\gamma Z}_2\}\over 2m^2_H(s-m^2_Z)}
\Bigg ] ~,  \label{V7-form}
\eqa
leading to a $\sin^2\theta\cos2\phi$ angular dependence. Such a dependence should
simply confirm the property of the $W_2$ structure functions, which of course could
also be obtained from $V_2$, but with a different $\gamma,Z$ combination.\\

\noindent
 Concerning (\ref{sigma1}-\ref{V7-form}), we add the following remarks:
\begin{itemize}

\item
If  $X$ consists of two scalar or  two vector particles, then  $W^{\alpha\beta}_3=0$.
This can be seen from (\ref{Walphabeta-i-1}, \ref{I123-forms}) by remarking that the
 $V$ and $H$ couplings to scalar or vector particles, can never
generate a parity violating $I_3$ contribution.\\

\item
In addition, at high energy and $p_H$, in the scalar cases $\gamma,Z \to s+s'$ followed
by $s\to H+s^{\prime \prime}$ or $s'\to H+s^{\prime \prime}$,
 the couplings are proportional to $p_s-p_{s'}$, so
 that only an $I_2$ contribution survives
  leading to $W^{\alpha\beta}_1\approx 0$.

Correspondingly, in the fermion case $\gamma,Z \to f+\bar f$ followed
by $f\to H+f$ or $\bar f\to H+\bar f$, the so-called
longitudinal part ($q^{\mu}q^{\nu}$)  also vanishes when masses can be neglected in
the kinematics, leading to
$W^{\alpha\beta}_1 + (p^2_H/m^2_H) W^{\alpha\beta}_2\approx 0$.

These properties should be useful  for identifying  the nature
of the $X$ states associated with  the Higgs boson.

\item
We recall that:
\bqa
 V_{1}, ~V_{4}  ~~ & {\rm  are ~  combinations ~ of }& ~~
W_1^{\gamma\gamma},~W_1^{ZZ},~W_1^{Z\gamma} ~ ,  \nonumber \\
 V_{2},~V_{5},~V_{7} ~~ & {\rm  ~ combinations ~ of}& ~~
 W_2^{\gamma\gamma},~W_2^{ZZ},~W_2^{Z\gamma} ~, \nonumber  \\
 V_{3}, ~V_{6}~~  & {\rm   ~combinations ~ of }&~~
W_3^{ZZ}, ~W_3^{Z\gamma} ~. \nonumber
\eqa

\end{itemize}

\section{The $H$+2-body contributions in  SM}

{\bf $s$-channel  $X$-forms in SM}\\
We first concentrate on the $s$-channel  $X$-forms that arise in SM through
the process (\ref{BSM-X}).
These are
\bq
X= t\bar t ~~,~~ f\bar f ~~,~~ZH~~,~~\label{s-channelX}
\eq
where $f$ contains all leptons and quarks except the $t$-one.
For such $X$-forms the general treatment
in Sections 2 and 3 fully applies.\\

For each final state we compute the standard amplitude. By taking its square
one obtains the $H^{\mu\nu}_{\alpha\beta}$ probability defined in Section 2.
In the $t\bar t$ case the amplitude is due to 3 diagrams
\[
e^-e^+\to \gamma,Z \to  t+(\bar t\to \bar t +H),~
e^-e^+\to \gamma,Z \to  \bar t+(t\to t +H),~
e^-e^+\to Z \to  H+(Z\to t\bar t)) ~.
\]
Similar diagrams occur for the light $f\bar f$ cases, but in this case
the first 2 ones are negligible due to the mass suppressed $Hf\bar f$ coupling.

In the case of $ZH$ one has 4 diagrams because the complete amplitude
producing the final $ZHH$ is symmetrized by exchange of the two Higgs-particles,
before computing the probabilities and the structure functions-particles
\[
e^-e^+ \to Z \to  H(p_H)+(Z\to Z+H(p'))~,~ e^-e^+\to Z \to H(p')+(Z\to Z+H(p_H))~,
\]
and the other (already symmetric) ones
\[
e^-e^+\to Z \to  Z+(H\to H(p_H)+H(p')) ~,~
e^-e^+\to Z \to Z+H(p_H)+H(p') ~,~
\]
involving the 4 leg $ZZHH$ coupling.

\begin{figure}[h]
\vspace{-1cm}
\[
\epsfig{file=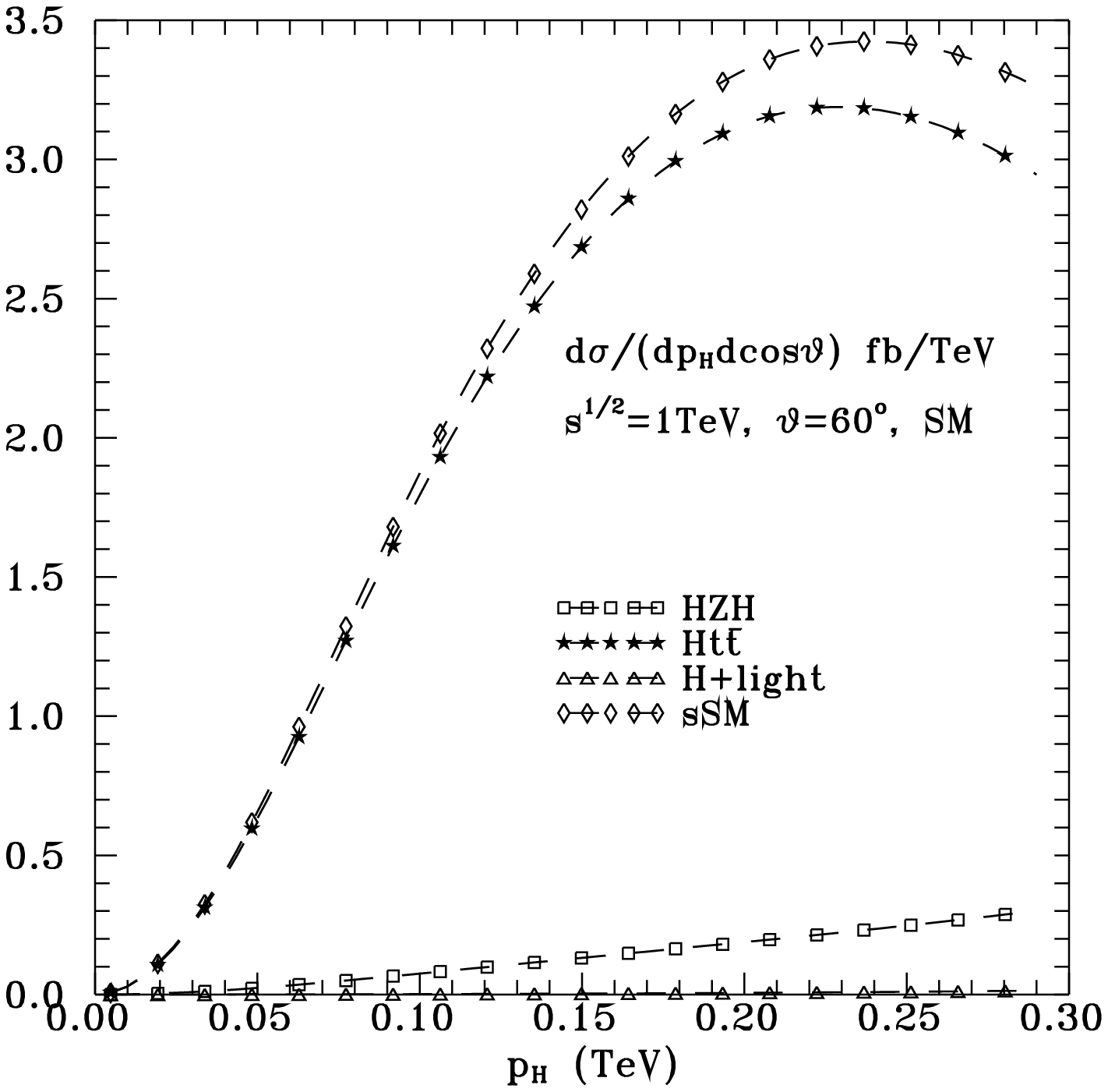, height=6.cm}\hspace{1.cm}
\epsfig{file=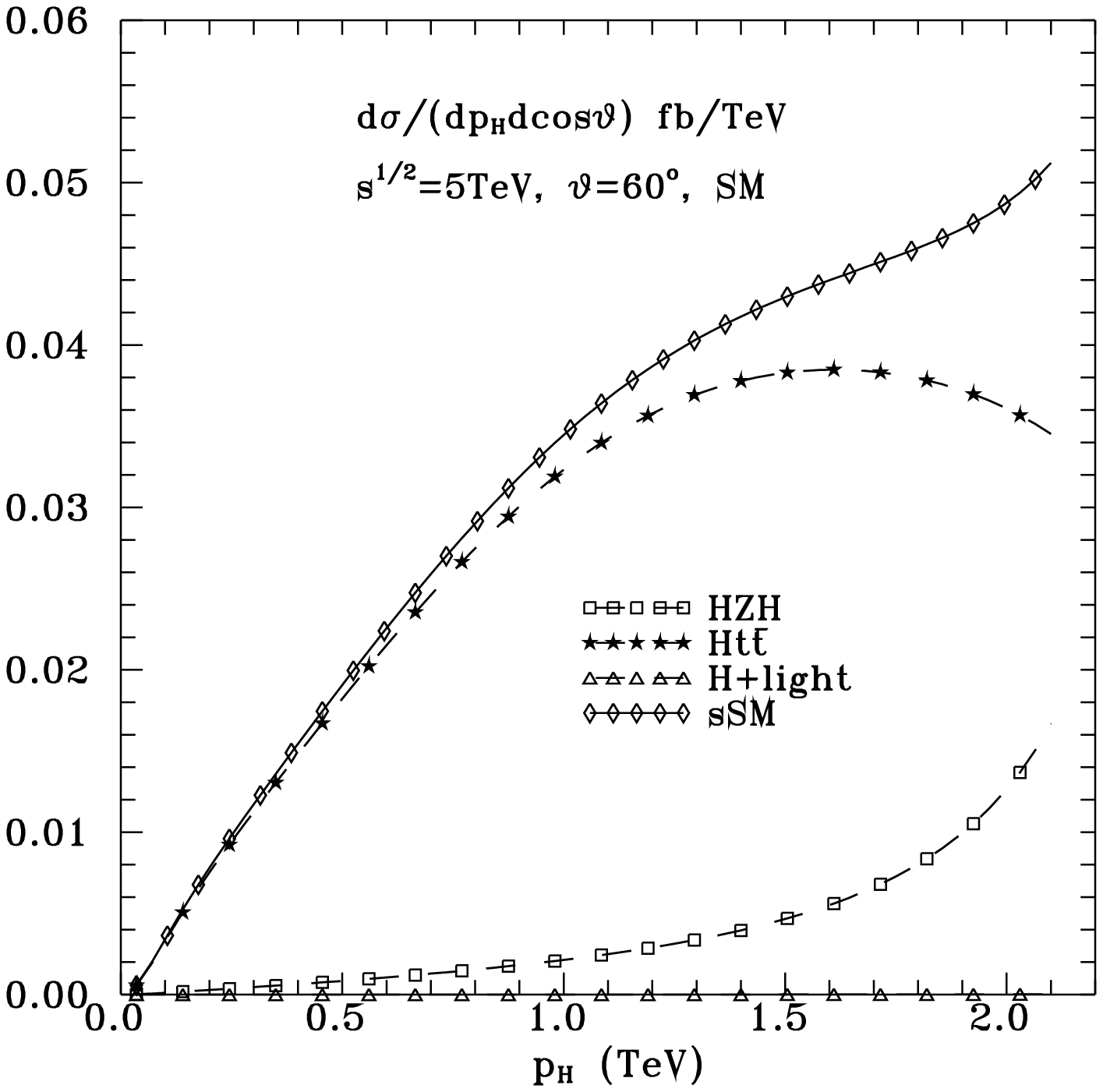,height=6.cm}
\]
\[
\epsfig{file=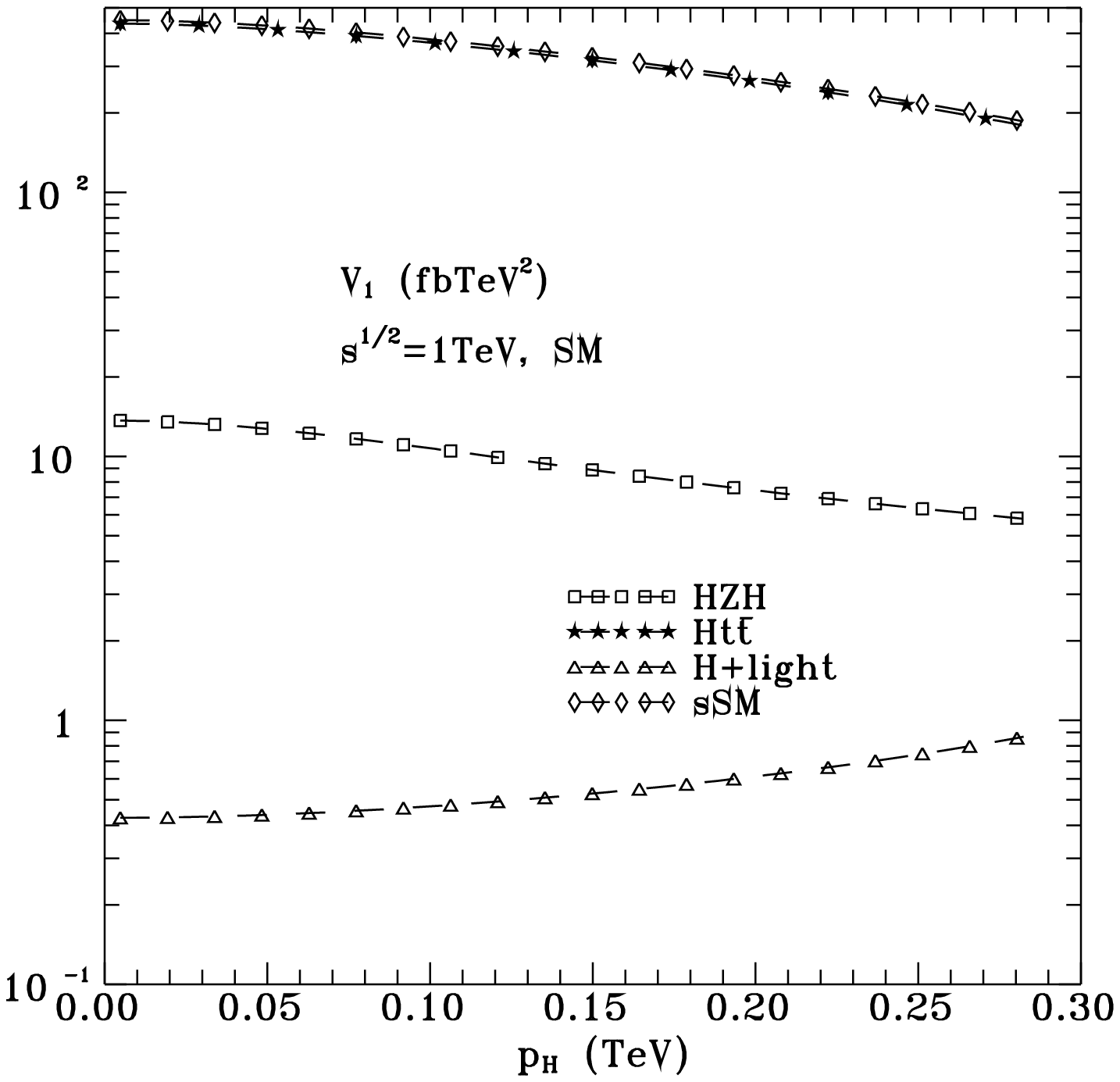, height=6.cm}\hspace{1.cm}
\epsfig{file=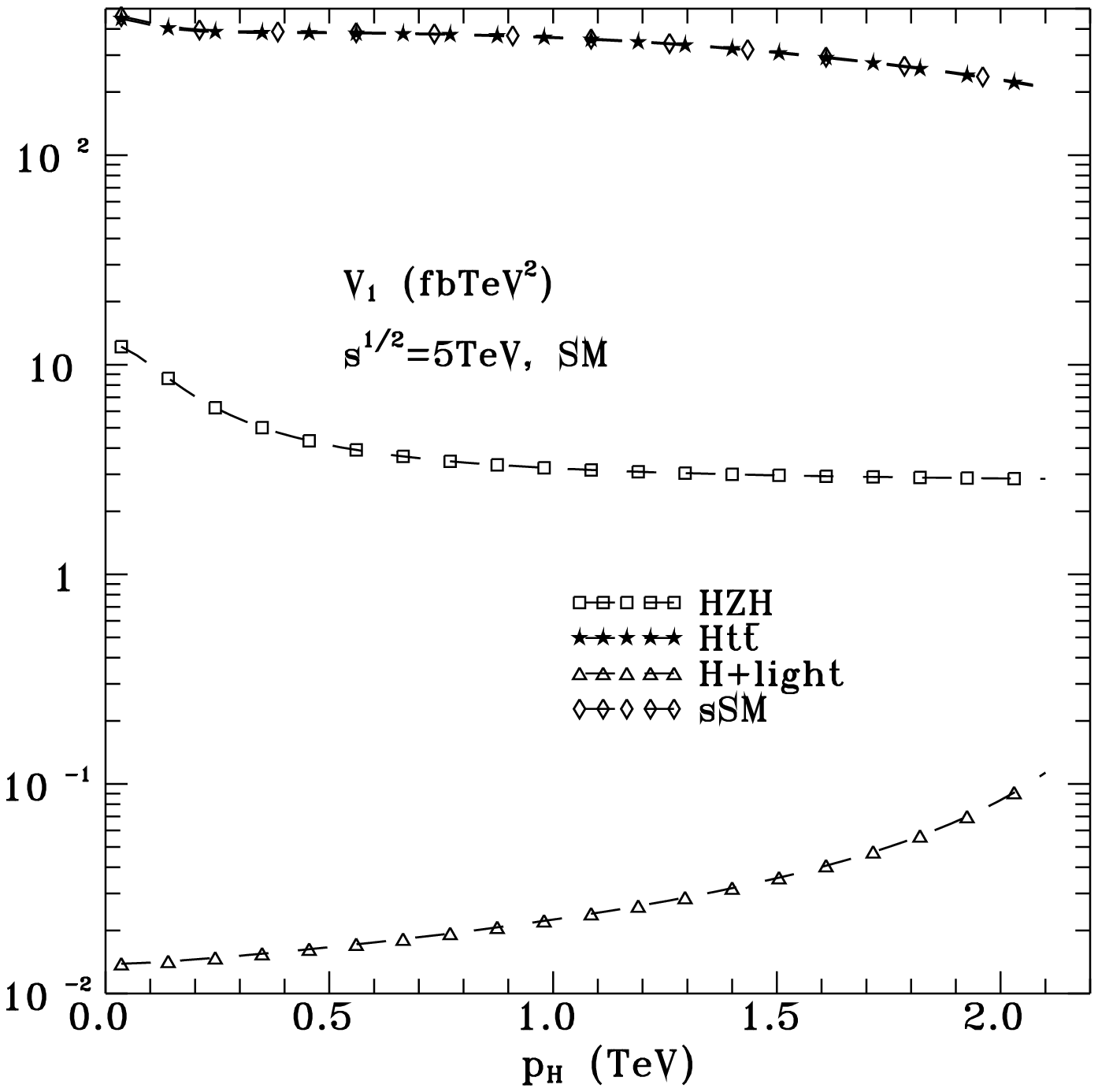,height=6.cm}
\]
\[
\epsfig{file=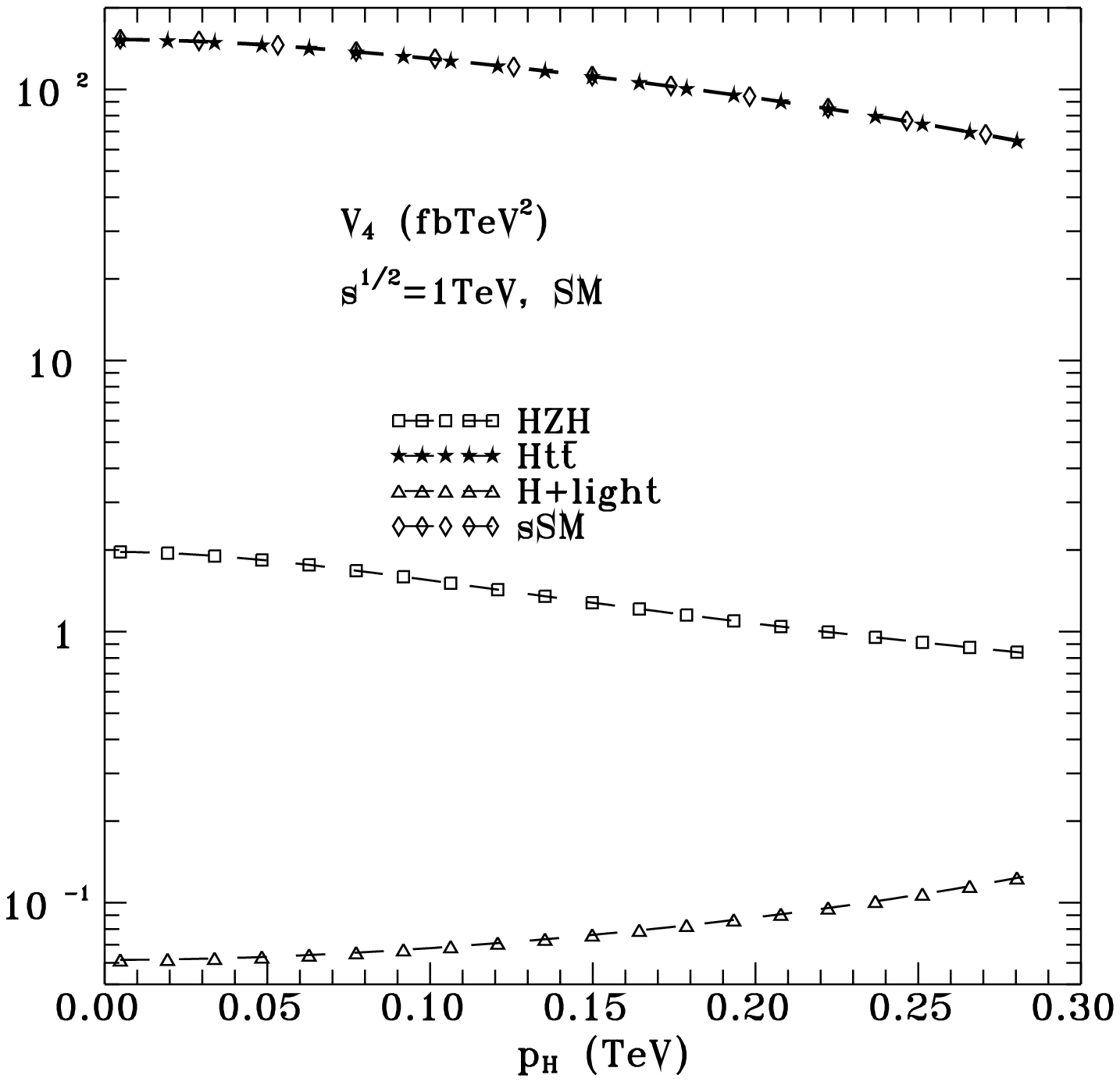, height=6.cm}\hspace{1.cm}
\epsfig{file=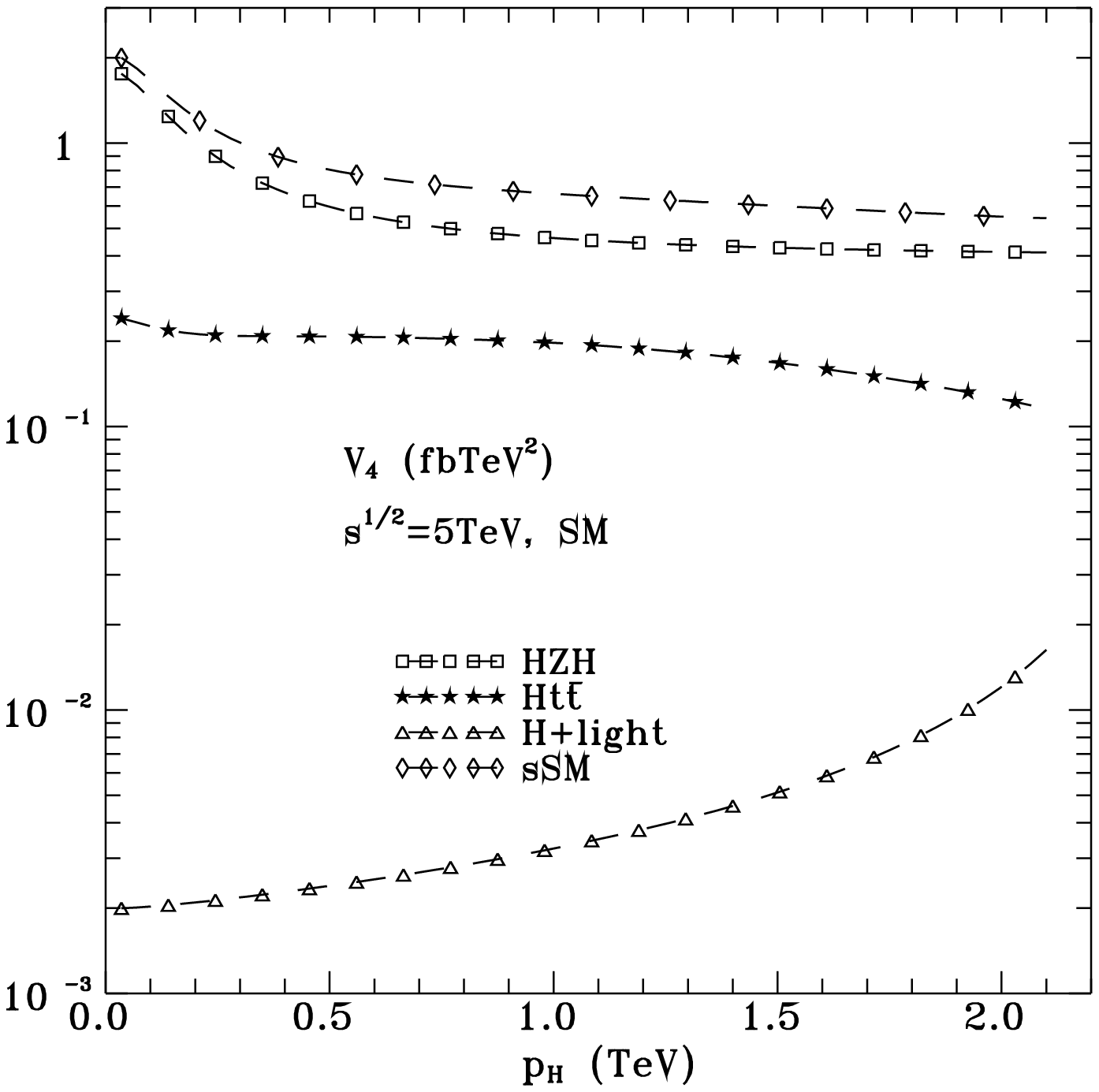,height=6.cm}
\]
\caption[1]{ Structure functions for the s-channel SM final states
  $X=ZH$, $X=t\bar t$,  and the lighter quark or lepton final states $X=f \bar f$
  called "light"; see (\ref{s-channelX}). Their   sum denoted as sSM is also given.
Left panels correspond to  $\sqrt{s}=1$TeV, and right panels to $\sqrt{s}=5$TeV.
Upper panels present the differential cross section measured in ${\rm fb/TeV}$, while
middle and lower panels denote respectively $V_1$ and $V_4$ measured in ${\rm fb TeV^2}$.}
\label{Gr1}
\end{figure}

\begin{figure}[h]
\vspace{-1cm}
\[
\epsfig{file=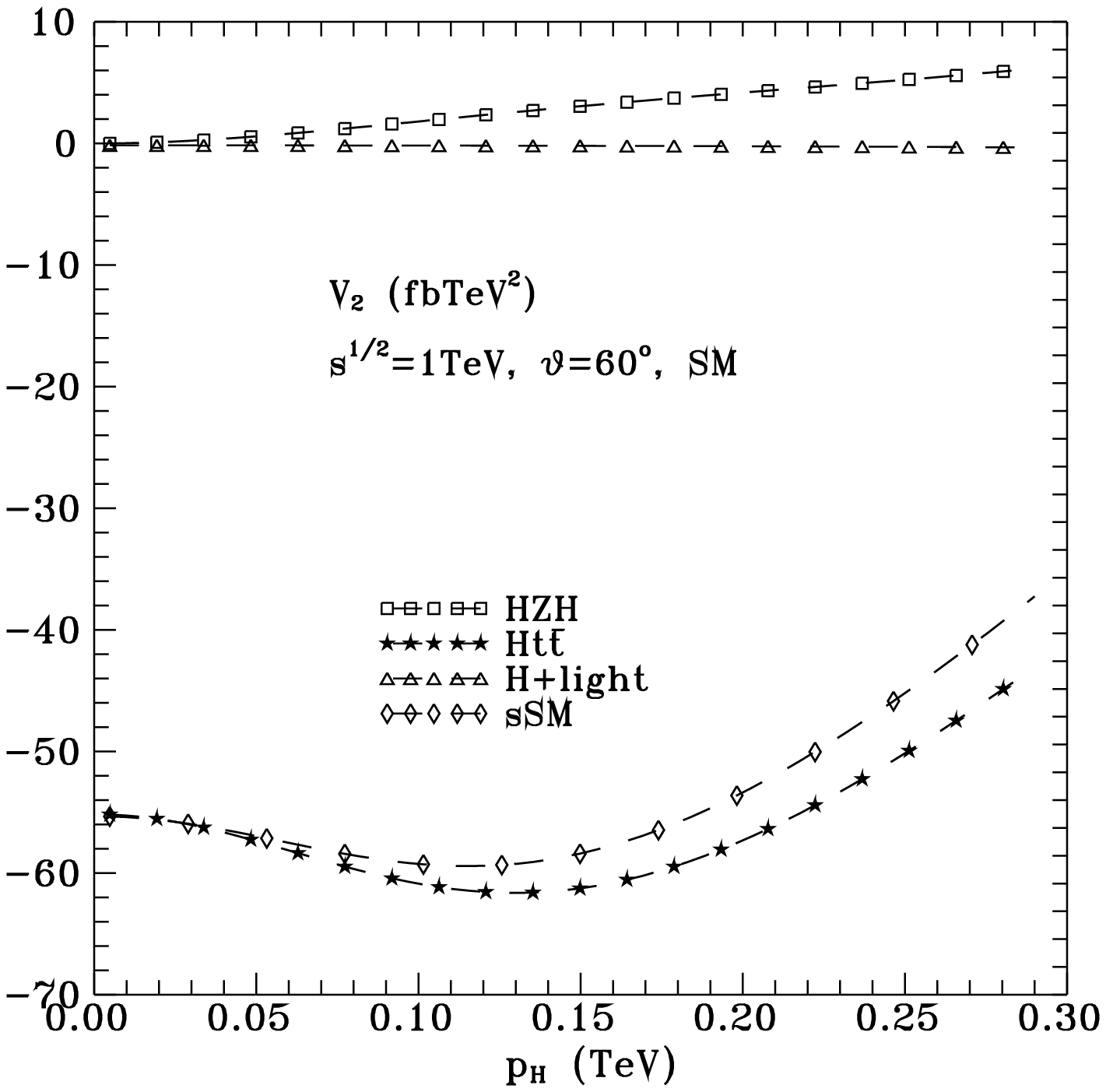, height=6.cm}\hspace{1.cm}
\epsfig{file=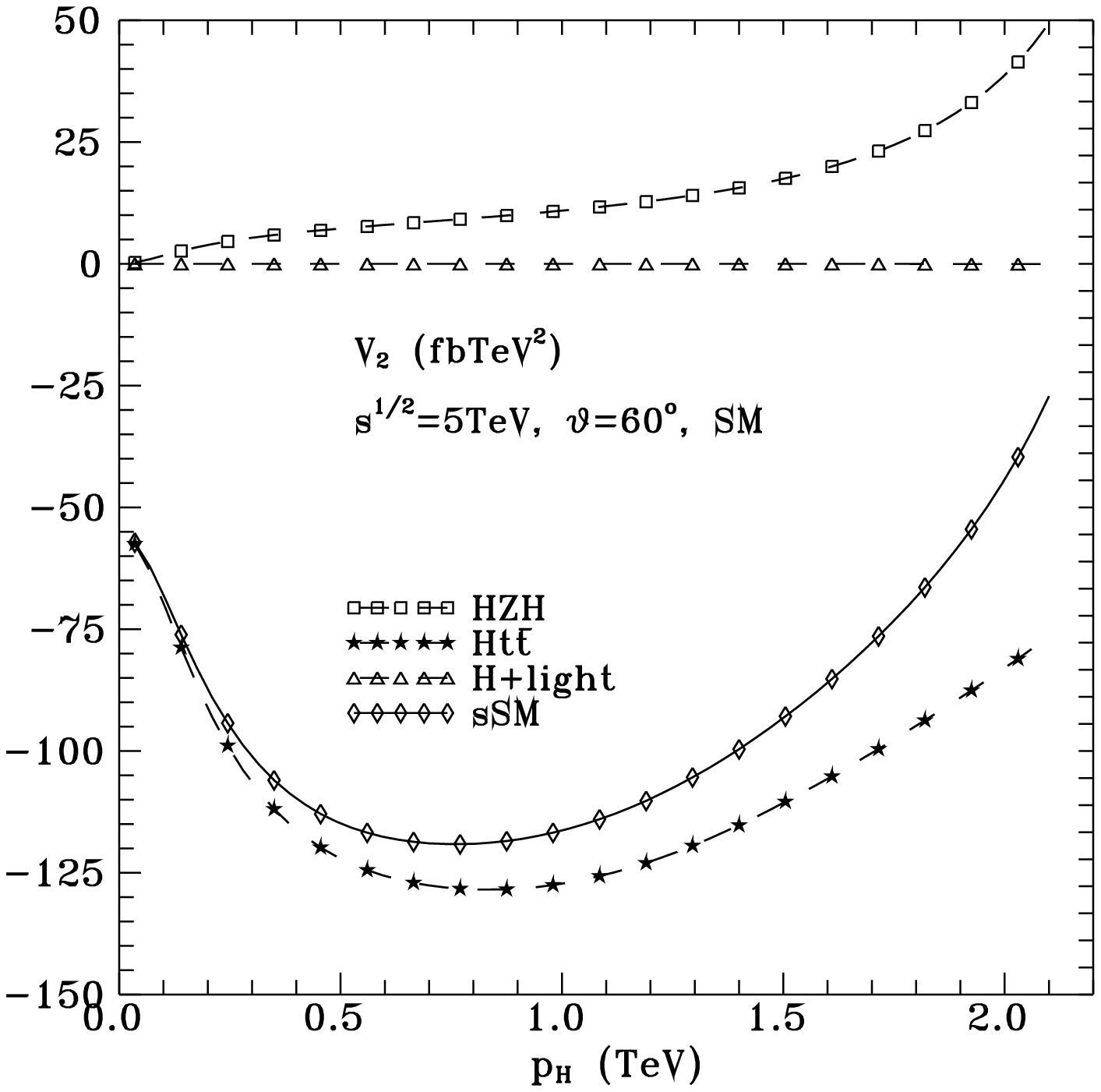,height=6.cm}
\]
\[
\epsfig{file=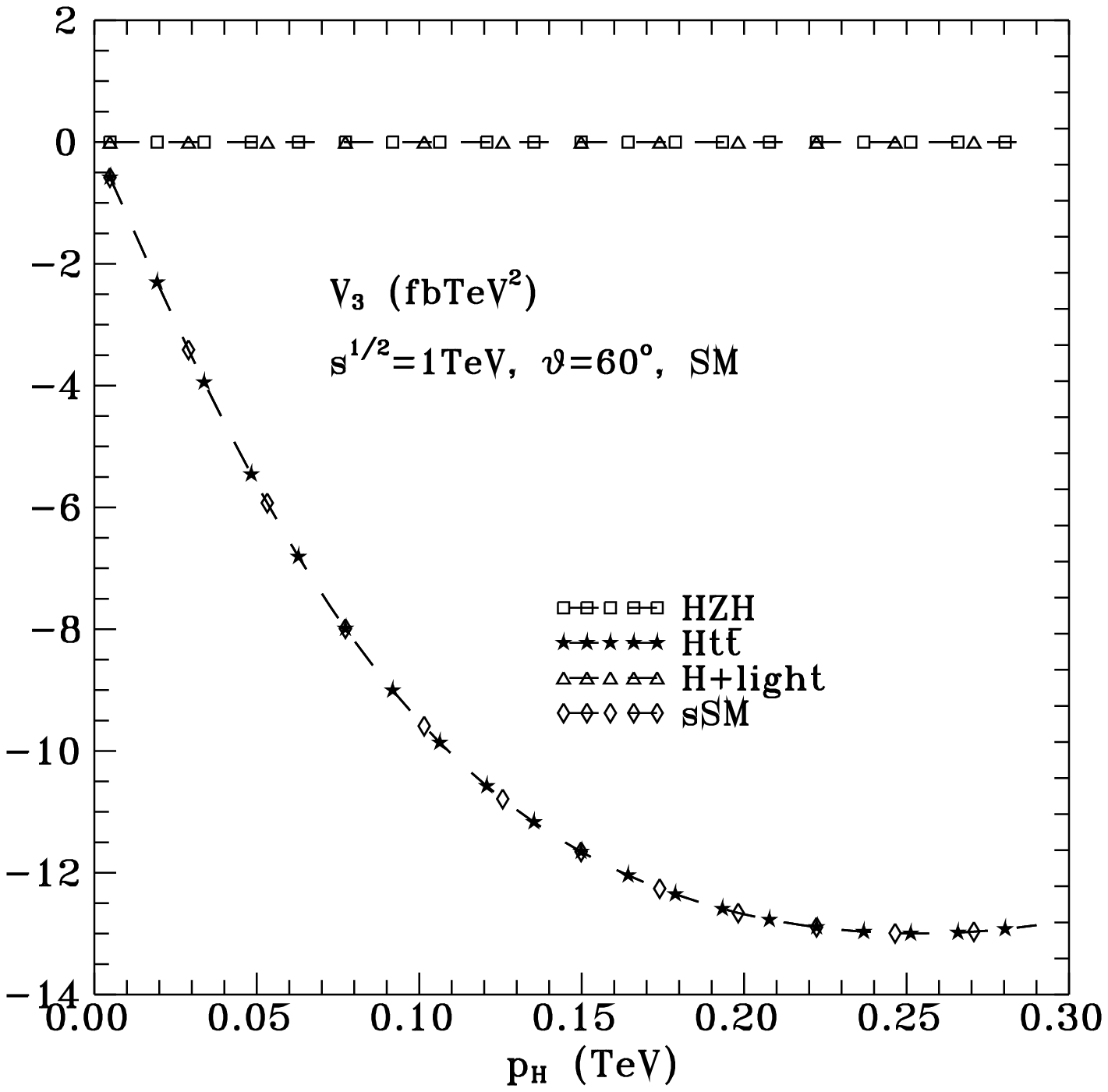, height=6.cm}\hspace{1.cm}
\epsfig{file=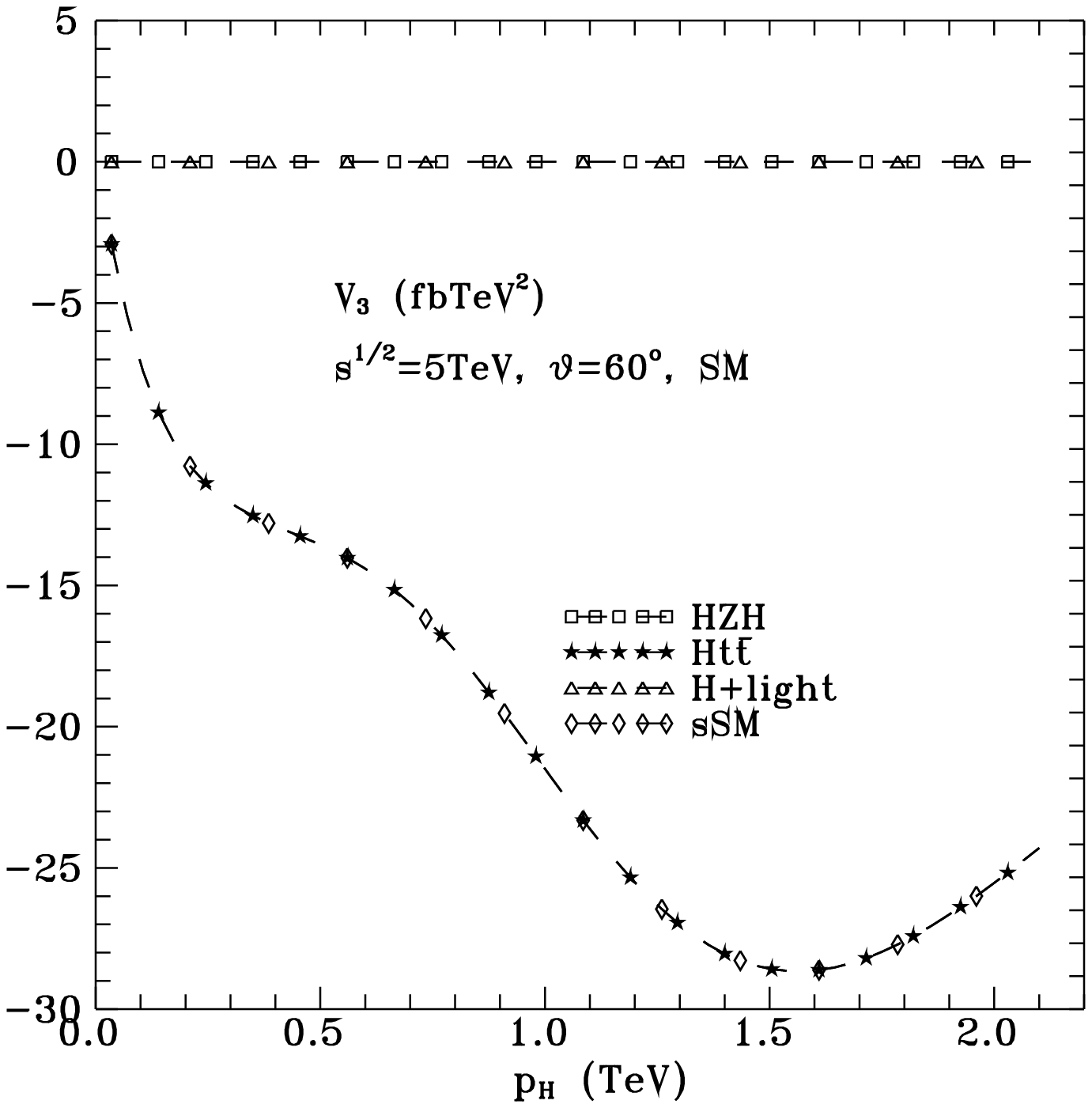,height=6.cm}
\]
\[
\epsfig{file=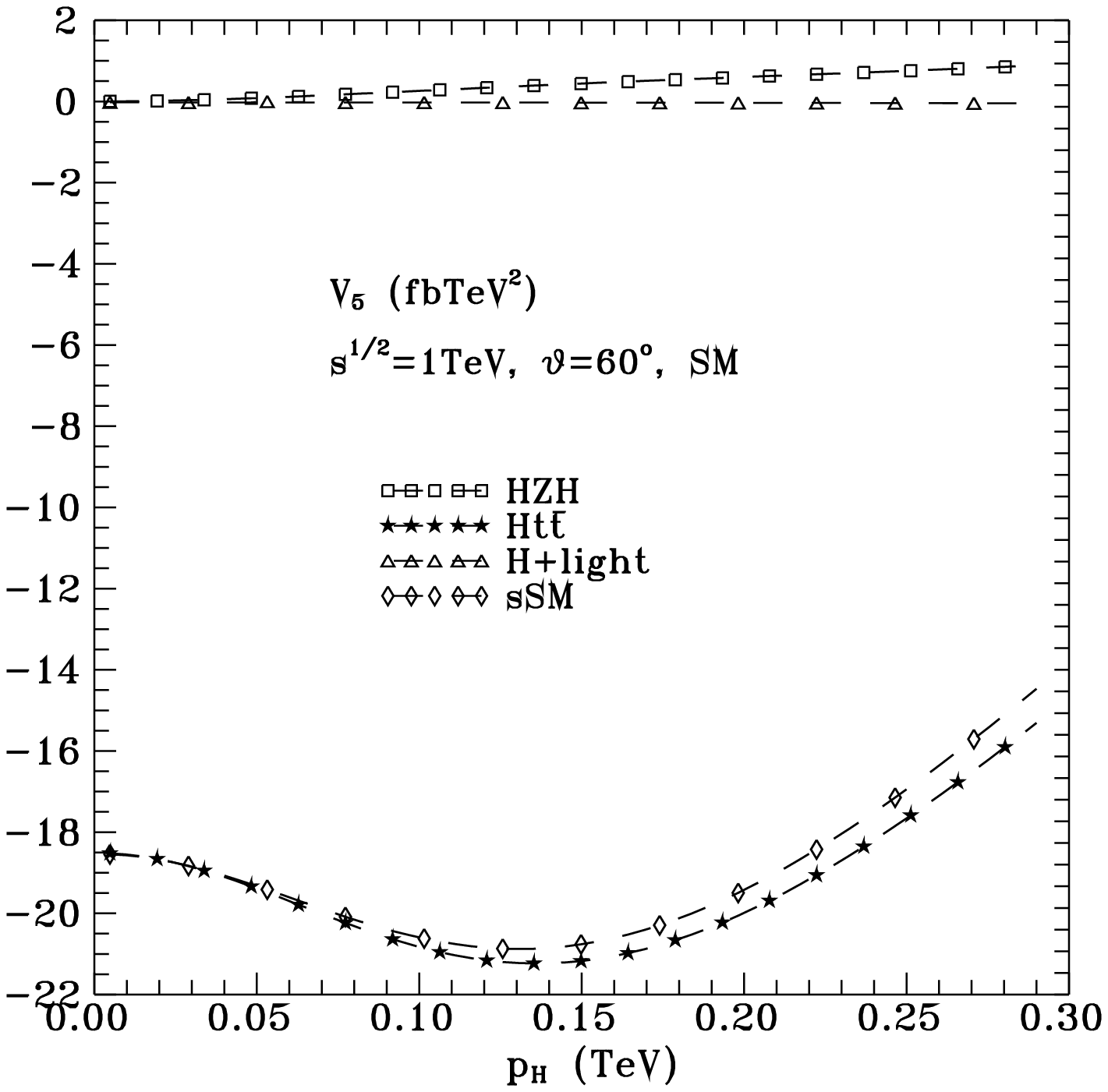, height=6.cm}\hspace{1.cm}
\epsfig{file=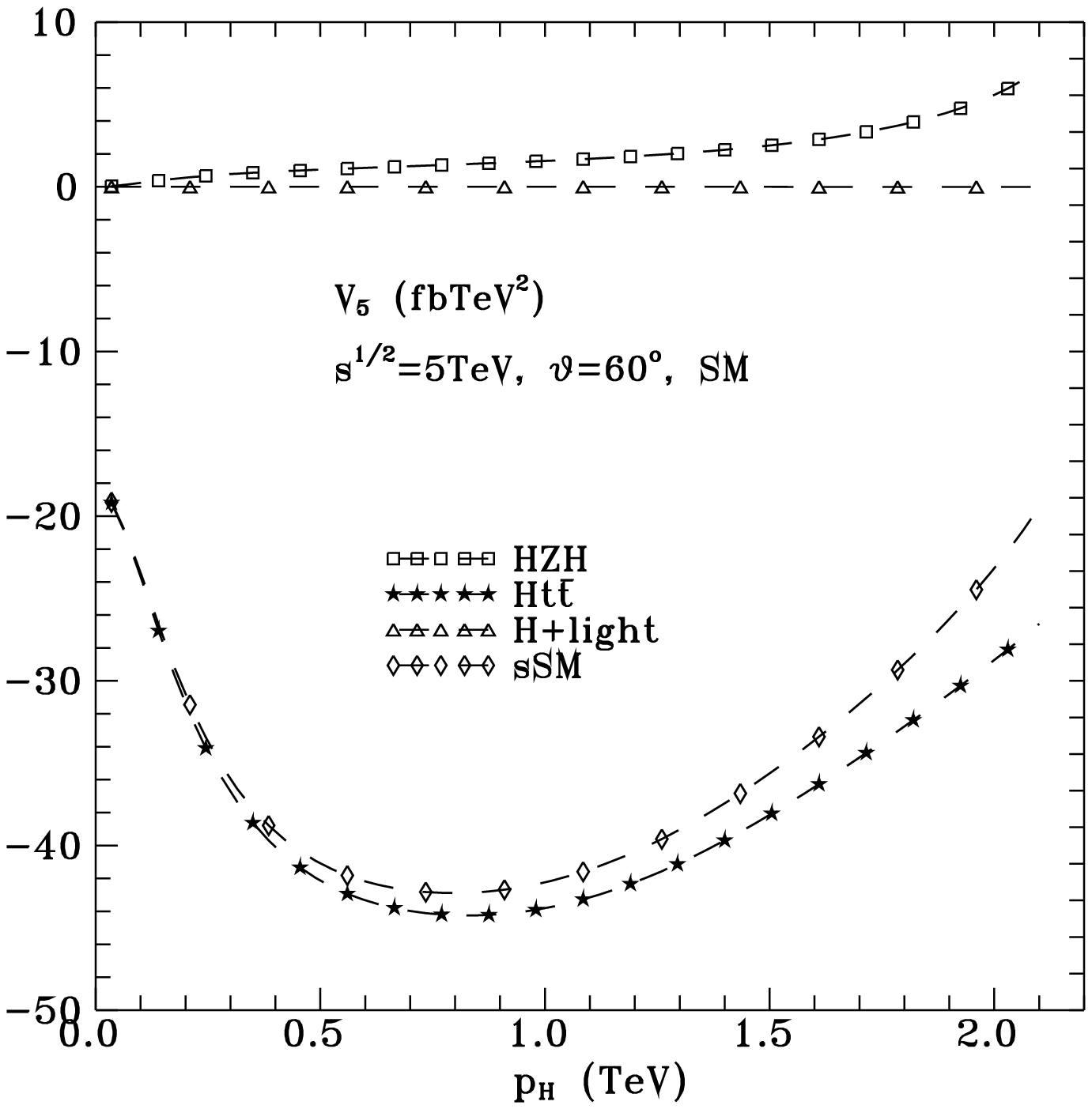,height=6.cm}
\]
\vspace{-0.5cm}
\caption[1]{ s-channel SM structure functions as in Fig.\ref{Gr1}. Upper, middle and lower panels
correspond  respectively to  $V_2,~V_3,~V_5$; see (\ref{V2-form}, \ref{V3-form}, \ref{V5-form}).}
\label{Gr2}
\end{figure}

\begin{figure}[t]
\vspace{-1cm}
\[
\epsfig{file=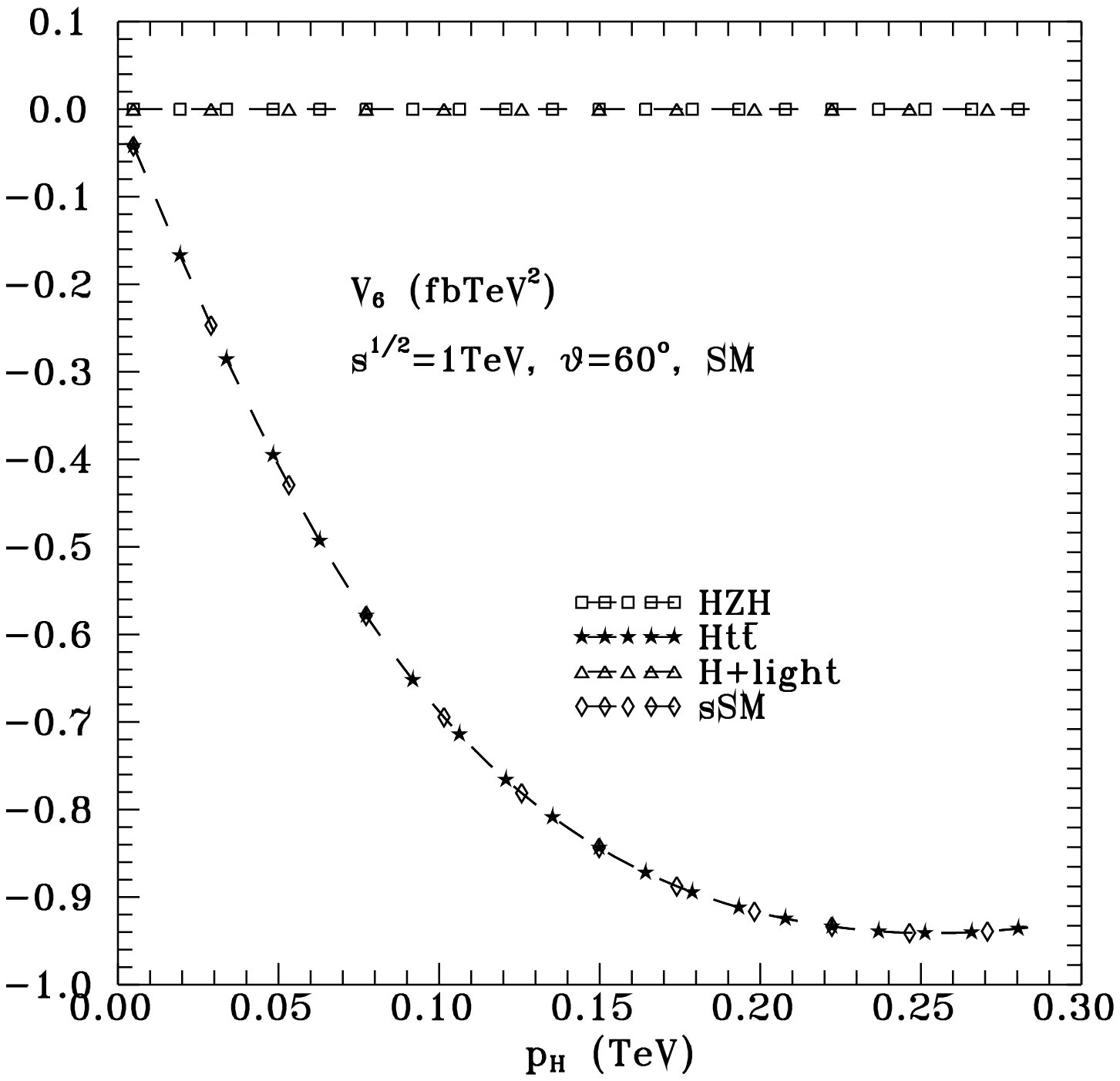, height=6.cm}\hspace{1.cm}
\epsfig{file=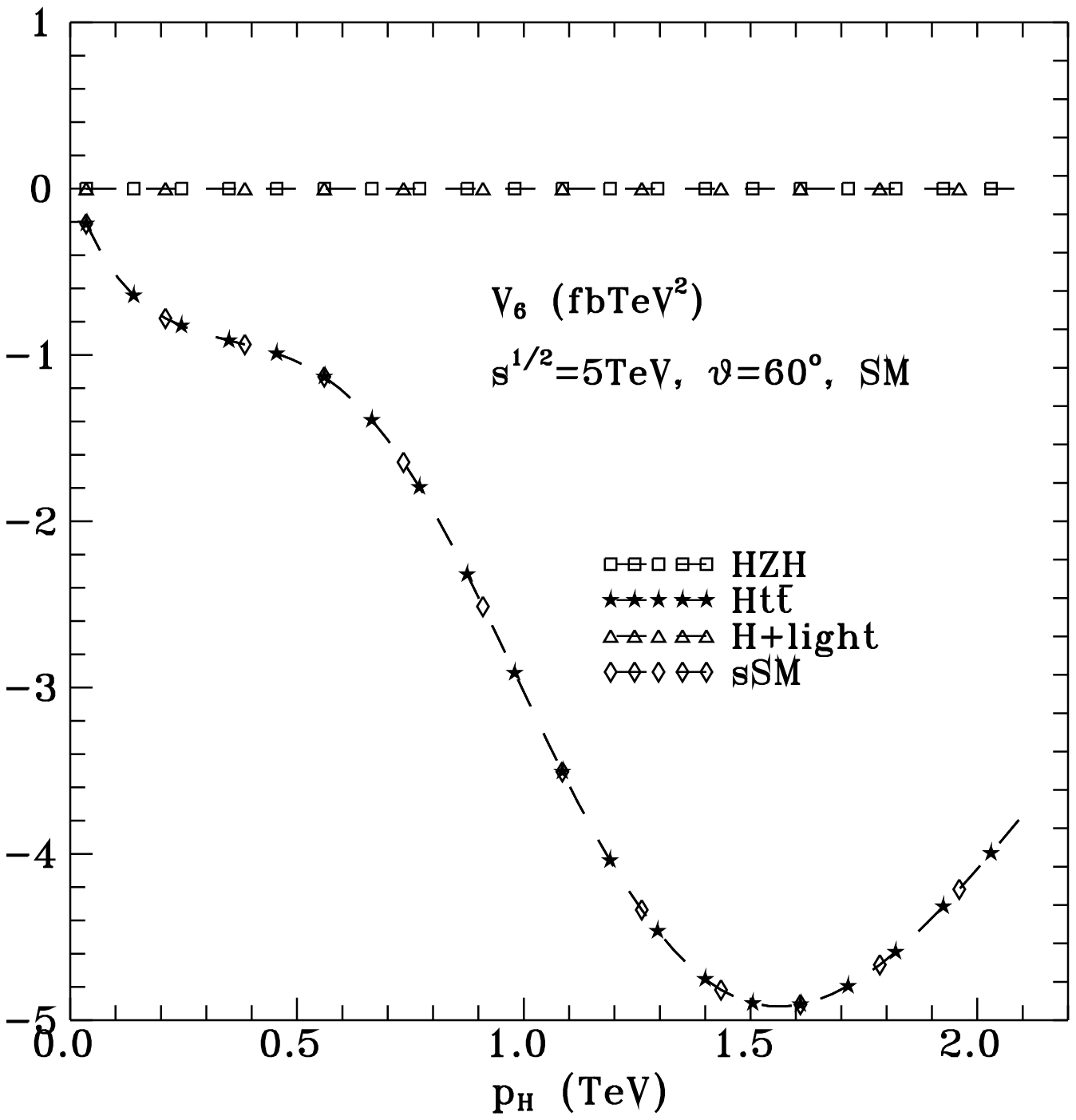,height=6.cm}
\]
\[
\epsfig{file=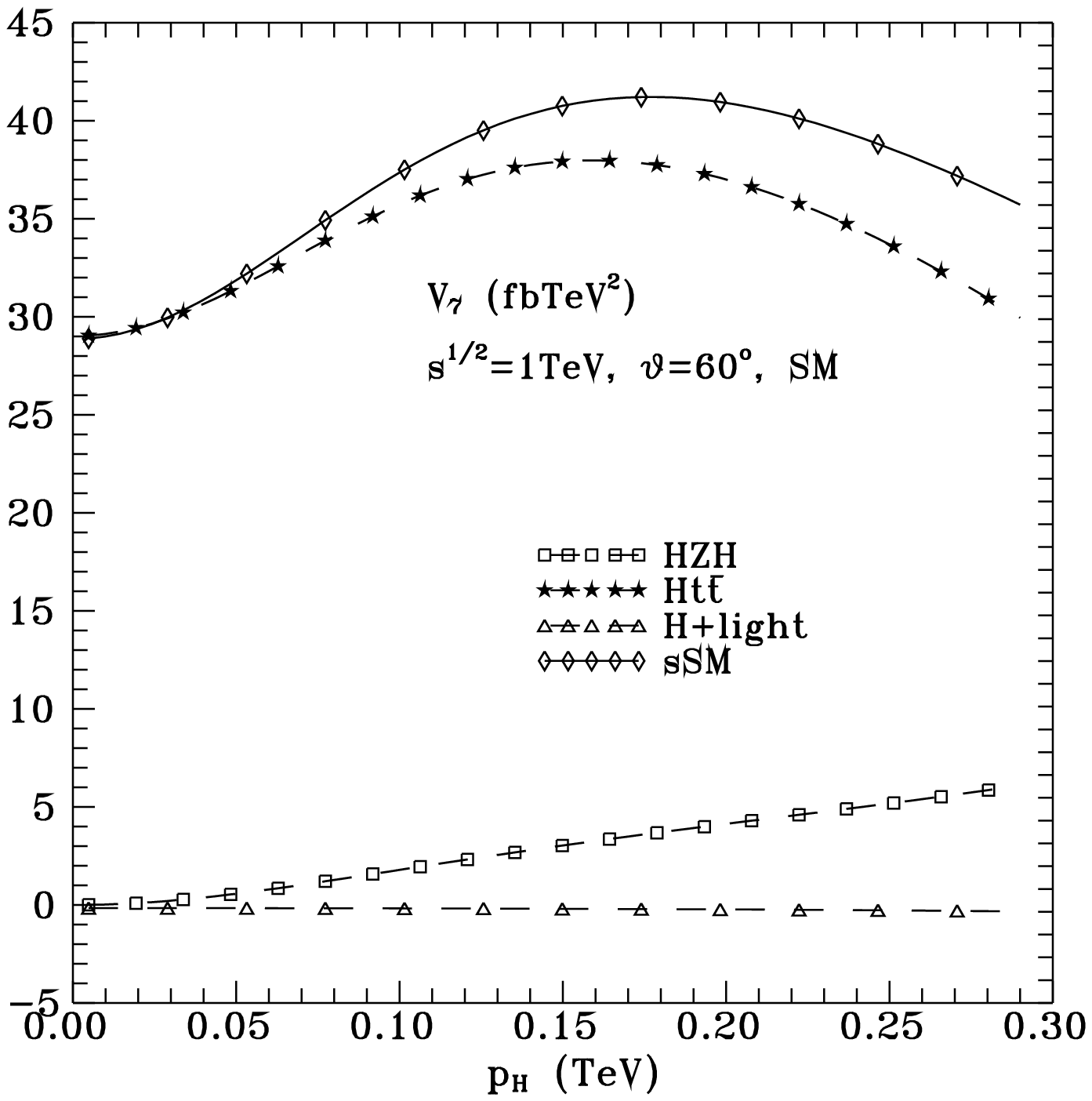, height=6.cm}\hspace{1.cm}
\epsfig{file=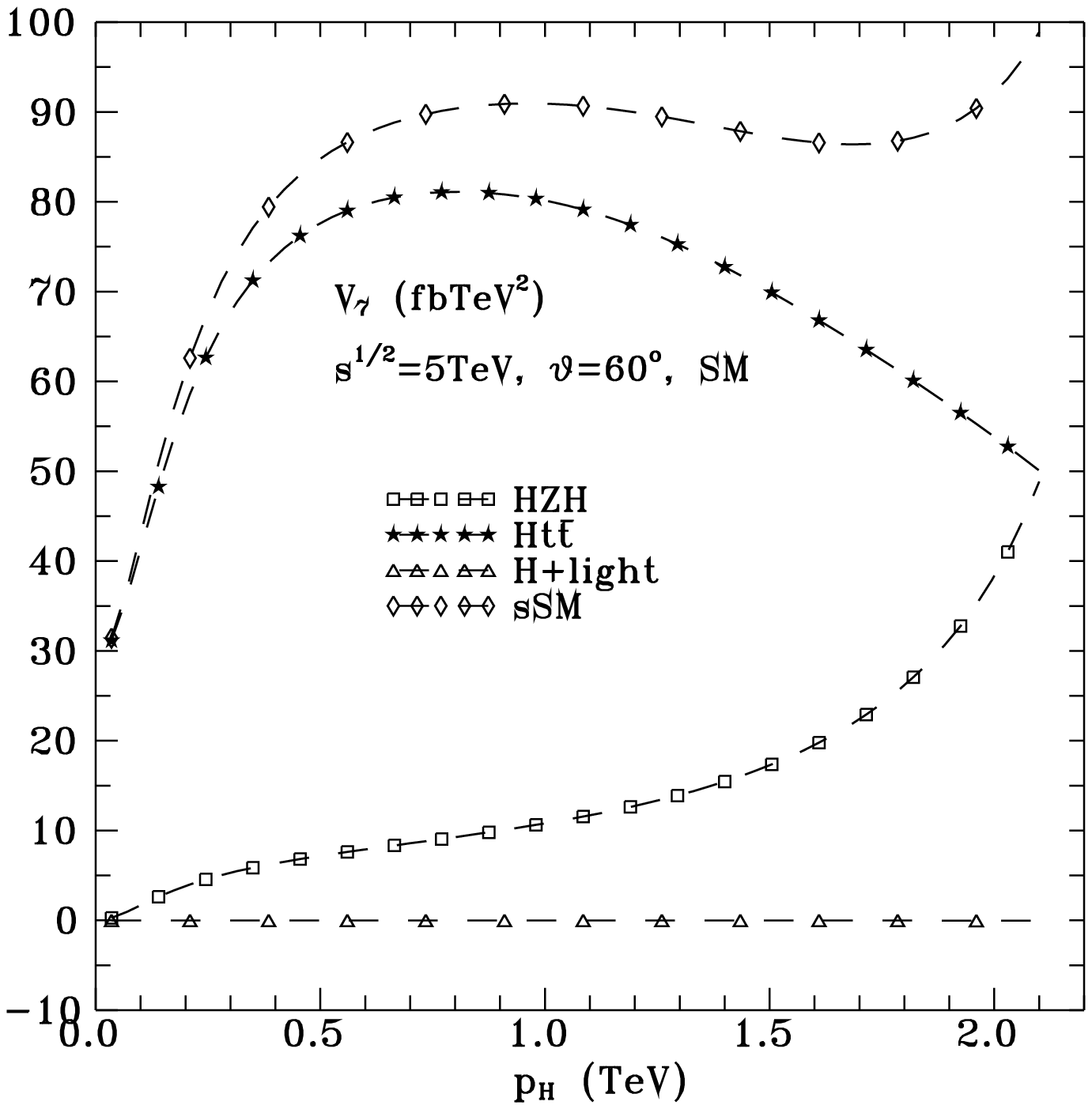,height=6.cm}
\]
\caption[1]{ s-channel SM structure functions as in Fig.\ref{Gr1}.  Upper and lower panels
correspond  respectively to  $V_6,~V_7$;  (\ref{V6-form}, \ref{V7-form}).}
\label{Gr3}
\end{figure}

We give in Figs.\ref{Gr1},\ref{Gr2},\ref{Gr3} the corresponding SM
unpolarized differential cross section (\ref{sigma2})
and the structure functions  $(V_1, ... ~V_7)$  appearing in
(\ref{V1-form}-\ref{V7-form}), for each of  the $X$ states in (\ref{s-channelX}).
In these figures, the $X=f\bar f$ state is always termed as "light". In addition to them, we also
give the results for the  sum of all the $X$-sates in (\ref{s-channelX}),
denoted as sSM.

In all Figs.\ref{Gr1}-\ref{Gr3}, we only give the dependencies on the Higgs momentum
 $p_H$, at a specific angle $\theta=60^\circ$; this angular specification is only dropped for
$(V_1,~V_4)$ which are angle-independent. Left panels always  correspond
to $\sqrt{s}=1$TeV, while right panels to
$\sqrt{s}=5$TeV.

In more detail,  Figs.\ref{Gr1} present the SM unpolarized differential cross section
and the angle-independent and usually largest $(V_1,~V_4)$. Figs.\ref{Gr2} show the angle-dependent
$V_2,~V_3,~V_5$, while Figs.\ref{Gr3} present $V_6,~V_7$.
In all cases, the $Htt$ contribution is the largest; then comes the $ZHH$ one, while the light fermions
contribution is much smaller.

Note also the differences in the shapes of these contributions,
as well as the specific connections among the $(V_1, ... ~V_7)$ mentioned in Section 3.
A possible violation of this, may supply hints on an  experimental departure
from the SM predictions, as we will see in the next section.\\

\noindent
{\bf $t$- and $u$-channel $X$-forms in SM}\\
In addition to the SM $s$-channel $X$ states appearing in (\ref{s-channelX}), one has to add the
$X$-states
\bq
X=W^-W^+ ~~,~~  ZZ ~~,~~ Z\gamma ~~,~~  \label{tu-channelX}
\eq
which differ by the presence of $t$ and $u$ channel neutrino or electron exchanges.

In the  case of the $H+W^-W^+$ final state, in addition to the 3 $s$-channel diagrams
\bqa
&& e^-e^+\to \gamma,Z \to  (W^-  \to W^- +H) + W^+ ~,~
e^-e^+\to \gamma,Z \to  W^- +(W^+ \to W^+ +H)~,~ \nonumber \\
&& e^-e^+\to Z \to  H+(Z\to W^+W^-), \nonumber
\eqa
one has the 2 t-channel neutrino exchange diagrams
\[
e^-e^+\to  (W^- \to W^- +H)+W^+ ~,~ e^-e^+\to  W^- +(W^+ \to W^+ +H)~.
\]
Finally, in the $ZZ$ and $Z\gamma$ cases one has only $t$ and $u$ electron exchanges
and one final $ZZH$ coupling.

\begin{figure}[t]
\vspace{-1cm}
\[
\epsfig{file=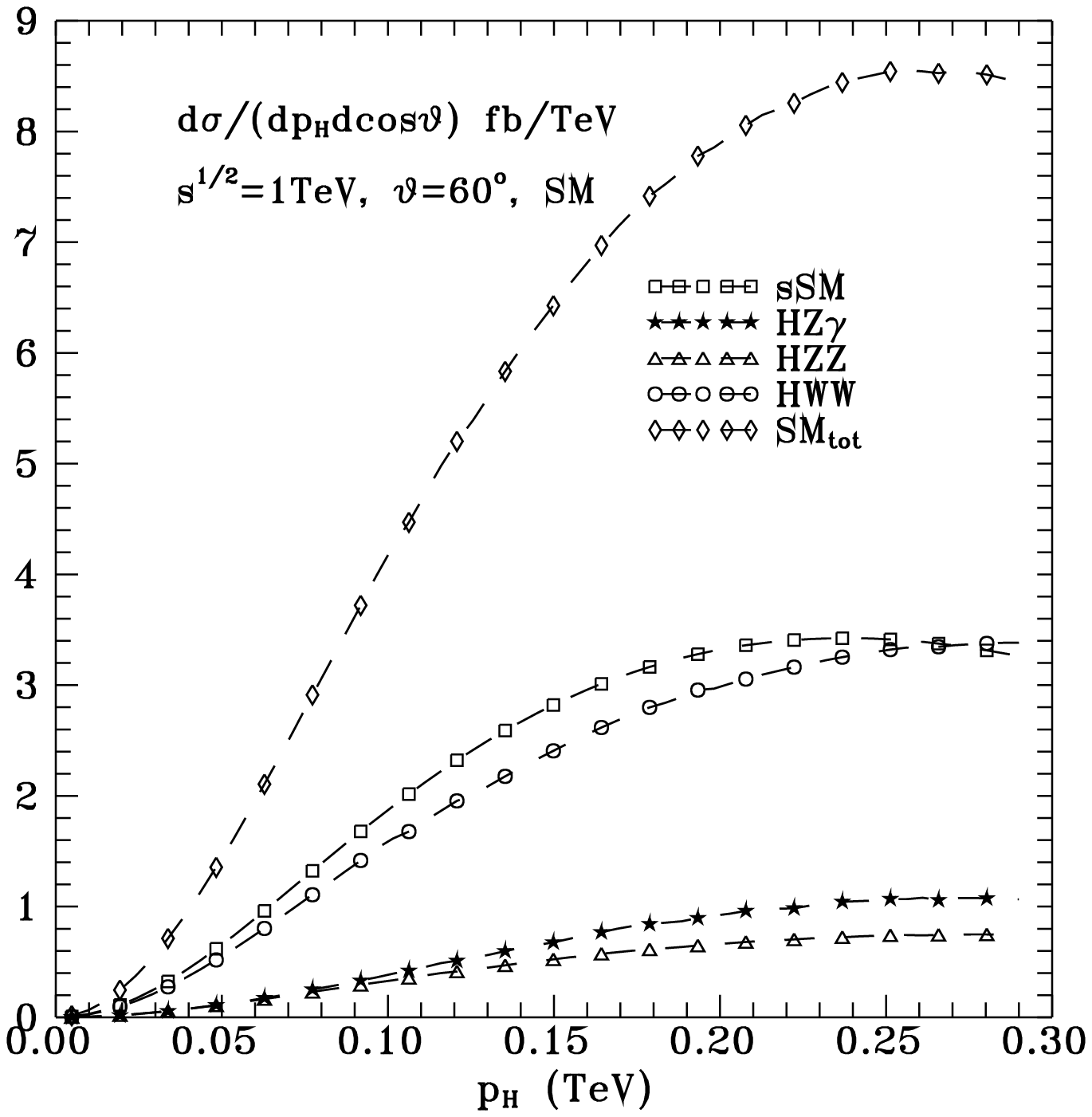, height=6.cm}\hspace{1.cm}
\epsfig{file=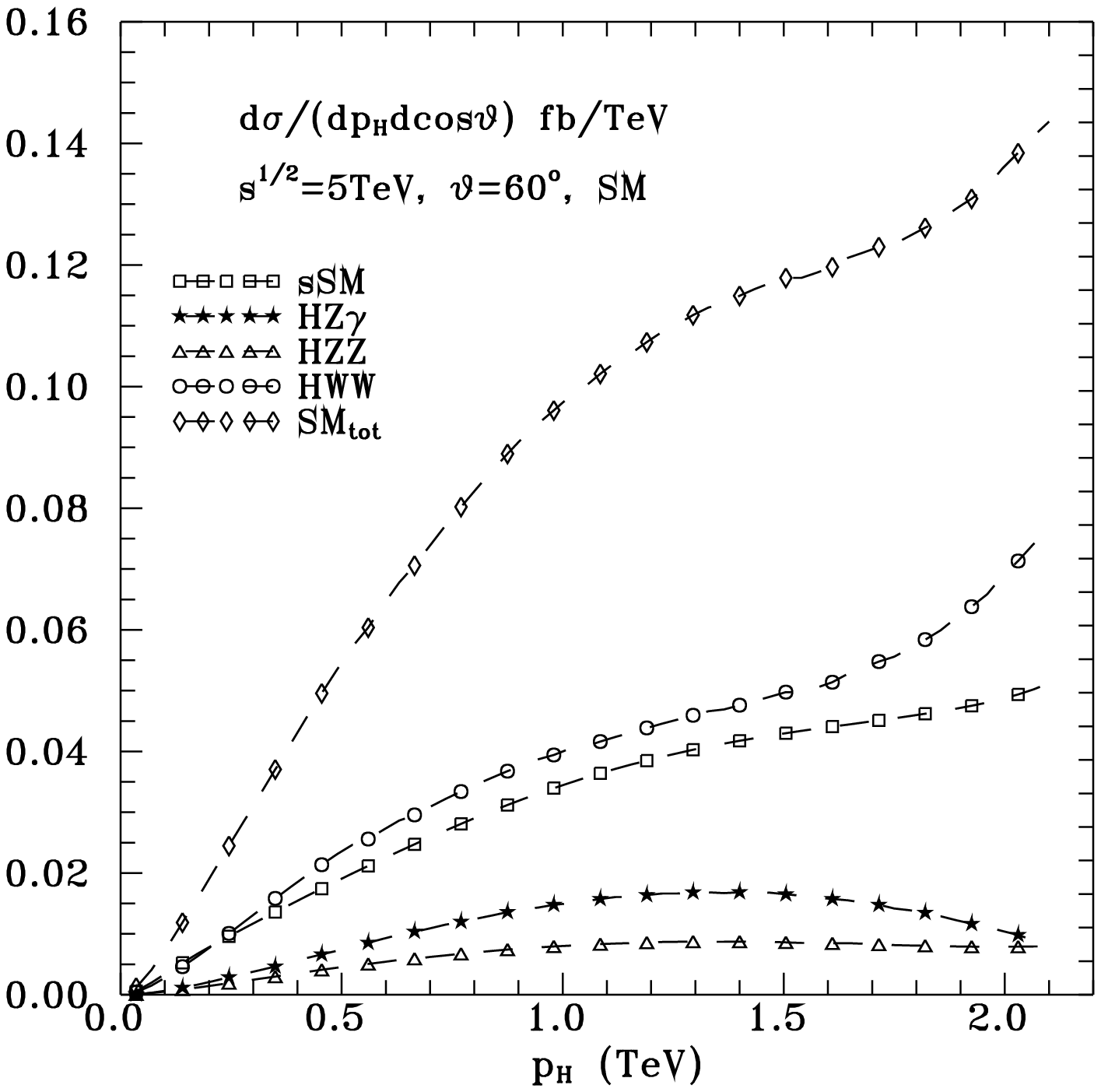,height=6.cm}
\]
\caption[1]{The SM differential cross sections (\ref{sigma2}) from the sum of the
the $X$-channels in (\ref{s-channelX}) called  sSM, together with the contributions
from the t- or u-channels in (\ref{tu-channelX}). The total contribution
from all channels (\ref{s-channelX},\ref{tu-channelX}) describes the total SM result
 called   ${\rm SM_{tot}}$. Left panel corresponds
to  $\sqrt{s}=1$TeV and the right one  to $\sqrt{s}=5$TeV. }
\label{Gr4}
\end{figure}

These $t$ and $u$ exchanges do not obey (\ref{BSM-X}) and thus, they cannot
be described  in terms of the simple Higgs boson structure functions (\ref{V1-form}-\ref{V7-form}).
Therefore,  we treat them separately and we directly compute their
contributions to the inclusive cross sections $d\sigma/ dp_Hd\cos\theta_H$. They are
illustrated   in Fig.\ref{Gr4}, as  functions of $p_H$.
In these illustrations we show  the contributions
of the sum of the  s-channel processes given in (\ref{s-channelX}) which has already appeared
in the upper panels of Fig.\ref{Gr1} (called sSM) and in addition, we also give
the contributions from the $X$-states in (\ref{tu-channelX}).
The sum  of all X-states in (\ref{s-channelX},\ref{tu-channelX}), which describes the total
SM contribution, called ${\rm SM_{tot}}$, is also given.\\

As already said, in order  to analyze experimental data in terms of the simple Higgs boson
structure functions (\ref{V1-form}-\ref{V7-form}), one has first to subtract
the contributions of the $X$-forms in (\ref{tu-channelX}) from the inclusive experimental
contribution to  $e^-e^+\to H+X$.

The same procedure should be
applied for the $ZZ$  and $W^-W^+$ fusion processes  $e^-e^+\to e^-e^+H$
and $e^-e^+\to \nu_e \bar\nu_e H$ (see \cite{ILC}). A priori these should be added
to the s-channel processes  $e^-e^+\to H+(Z \to e^-e^+,\nu_e \bar\nu_e )$. But, since their
contribution to the H structure functions is found to be negligible,
we  ignore the interference of these 2 sets of amplitudes.
One can then subtract the theoretical expectations for the fusion processes
from the $e^-e^+ \to H + X$ experimental cross sections, and then analyze the data
according  to (\ref{sigma2}, \ref{V1-form}, ...\ref{V7-form}) leading to
Figs.\ref{Gr1},\ref{Gr2},\ref{Gr3}.\\

\section{New contributions}

We now turn to possible new physics contributions. These are chosen to obey (\ref{BSM-X}), so that
the description in terms of the $V_i$ in (\ref{V1-form}-\ref{V7-form}) remains possible.
We treat the following examples:

\begin{itemize}

\item
  {\bf Possible new scalar particles $s$}.
We compute the amplitudes of the process $e^+e^-\to V \to  ss$, followed by $s\to H+s$
(2 diagrams). If $s$ is neutral, then the only gauge boson
it can couple to is $Z$. If $s$ is charged, then it can also couple to the photon.
Suitable values for the $ssH$-couplings are chosen,
so that the overall magnitudes of  the new physics effects
are  comparable to the SM one. The discrimination of new physics effects is then mainly based
on  studying the shapes  of the cross section and the structure functions.

We consider three possible $s \bar s$-pairs, using different $s$-masses:\\
s1: $s$=neutral, $m_s=0.1$ TeV,\\
s2: $s$=neutral, $m_s=0.28$ TeV,\\
s3: $s$=charged, $m_s=0.28$ TeV.\\
In all cases, we find that  heavy masses give quickly a strong decrease at high $p_H$.

\item
 {\bf New  fermion particles $b'$}.
 The involved new physics is described by two diagrams through the processes
$e^-e^+\to V \to b'\bar{b'}$, where $V=\gamma,Z$, followed by
$b'\to H+b'$ or $\bar b'\to H+\bar b'$, where
$b'$ is a new heavy $b$ fermion. For the illustrations we choose
$m_{b'}=0.25$ TeV. This way, one may study how the distributions among the various
$V_i$ reflect the fermionic nature.

\item

{\bf A scalar parton model.}
Another type of a schematic example could be obtained by imitating the parton model of hadronic
Deep Inelastic Scattering. Starting e.g. by assuming that transitions like $\gamma,Z \to x+x'$ exist,
where $x$ and $x'$ are particles that may fragment to $H$, through
transitions like  $x\to H + \rm{``anything"}$ and $x'\to H + \rm{``anything"}$.
The parton model fragmentation functions $D^H_x(z)$ and
 $D^H_{x'}(z)$,  are modelized for example through  typical
$z^a(1-z)^b$ forms,  with $z=2p^0_H/ \sqrt{s}$.
 The corresponding $W^{\alpha\beta}_i$ structure functions are then obtained by multiplying
the basic $e^-e^+\to \gamma,Z \to x+x'$ cross section by the above fragmentation
functions.

Examples could be given by taking $x,x'$ as scalars, fermions or gauge bosons, and correspondingly
choosing their fragmentation functions and the corresponding $W^{\alpha\beta}_i$ ones.
Here we use a simple example with a neutral scalar $x=x'$ coupled to  $Z$ only.
Neglecting  masses for simplicity, we obtain a non-vanishing contribution
only for  the $W^{ZZ}_2$, given by\footnote{As usual, the normalization
$\int_0^1 z D(z) dz=1$ is used.}
\bq
W^{ZZ}_2=g^2_{Zeff}{8m^2_H\over sz^3}D(z)~~, ~~ {\rm with} ~~ D(z)=12z(1-z)~~. \label{parton1}
\eq
In such a case, non-vanishing ``parton" contributions are only generated for
$V_{2,5,7}$. The choice  $g^2_{Zeff}\simeq 0.04$ is made  in (\ref{parton1}), so that this
 contribution has a magnitude  comparable to the other new physics effects discussed above.

\end{itemize}

\begin{figure}[b]
\[
\epsfig{file=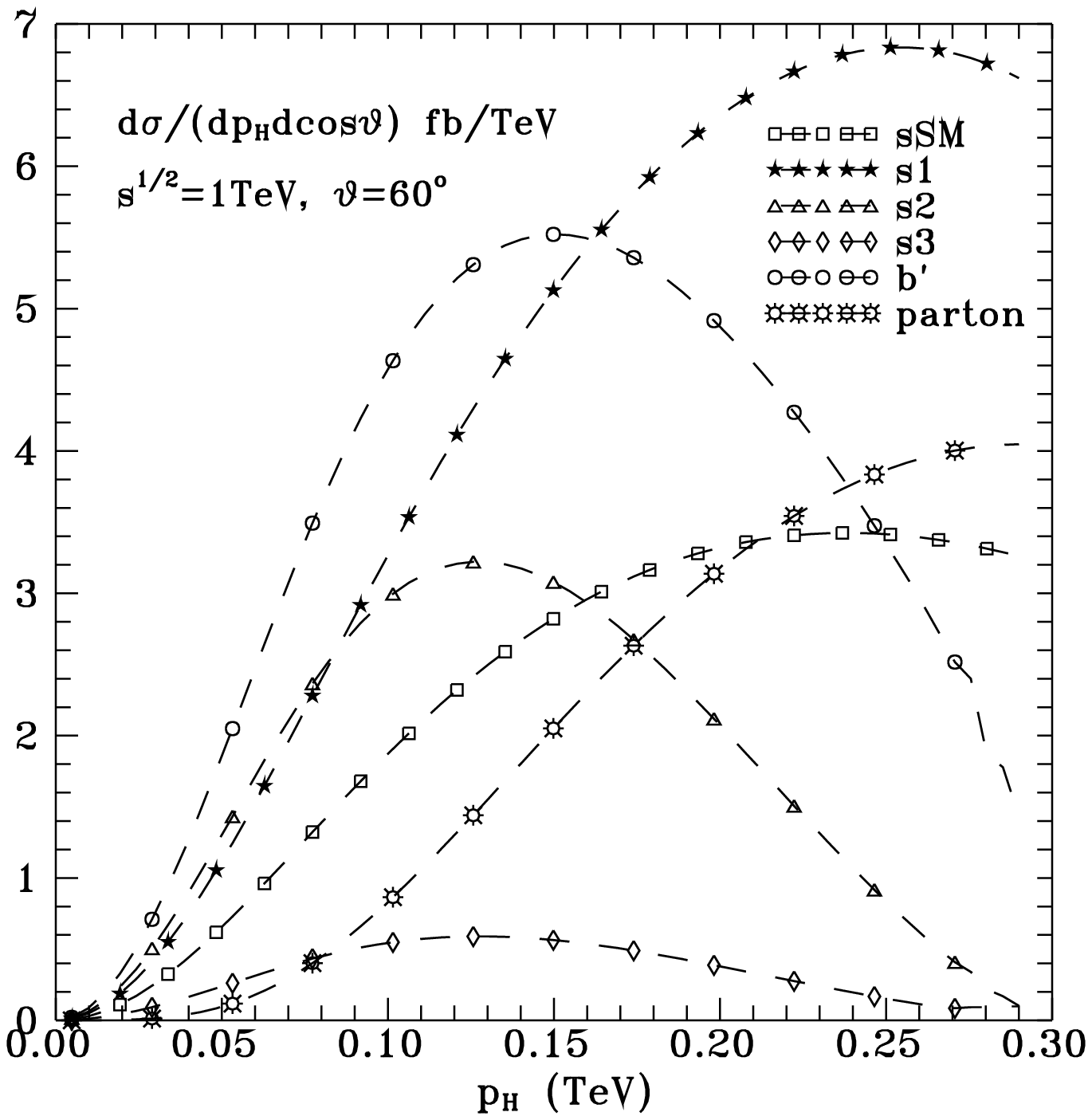, height=6.cm}\hspace{1.cm}
\epsfig{file=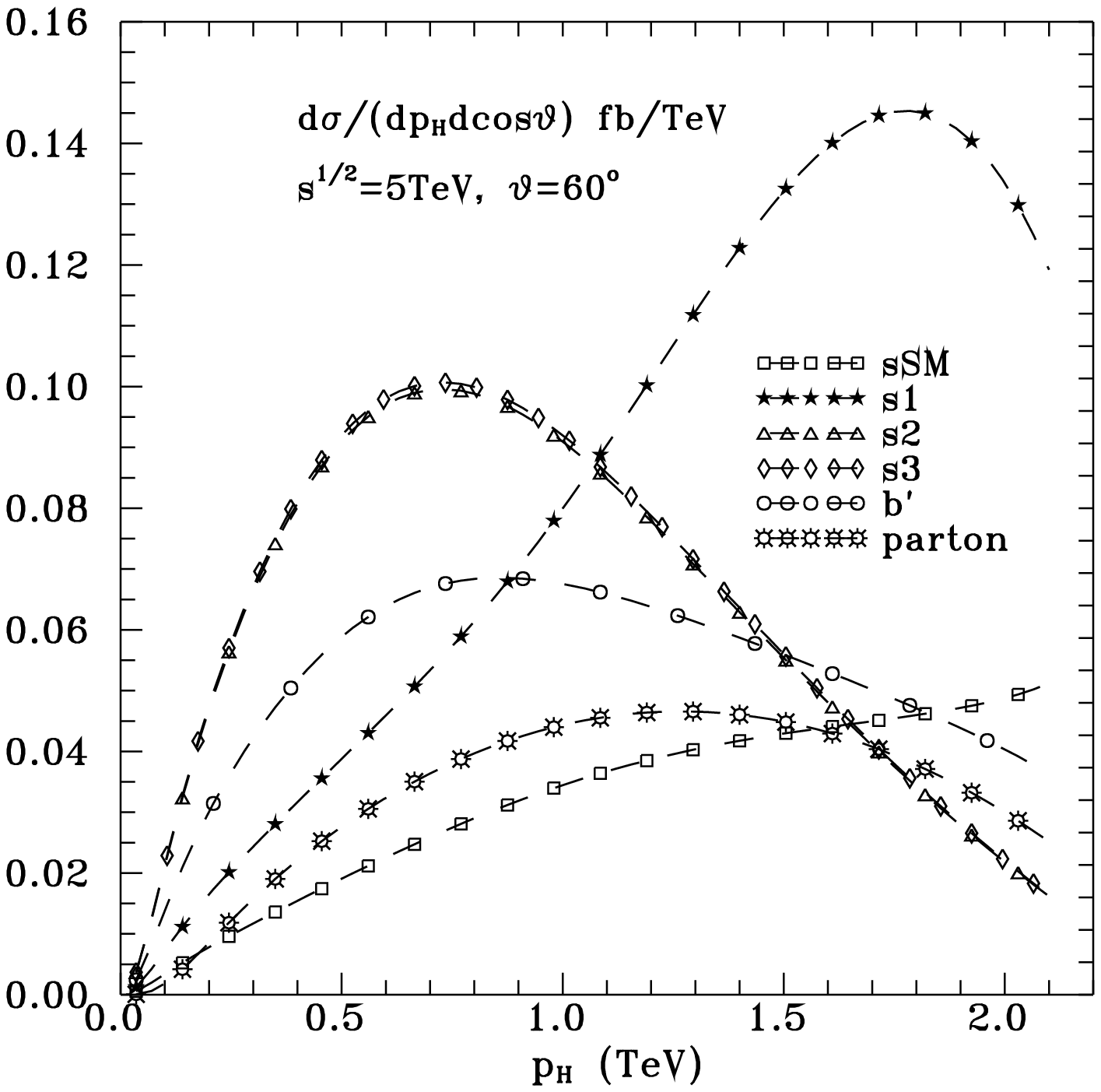,height=6.cm}
\]
\[
\epsfig{file=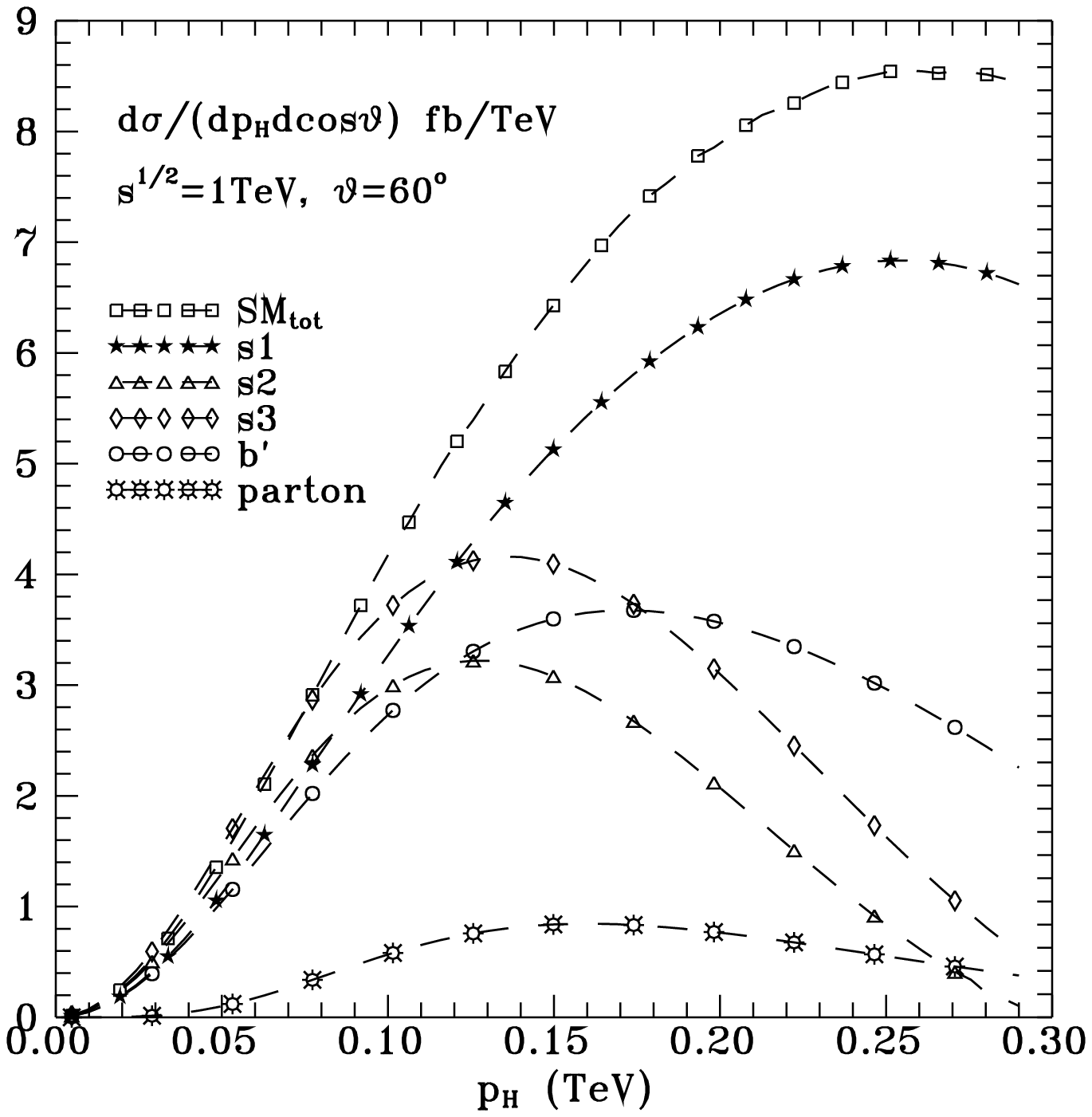, height=6.cm}\hspace{1.cm}
\epsfig{file=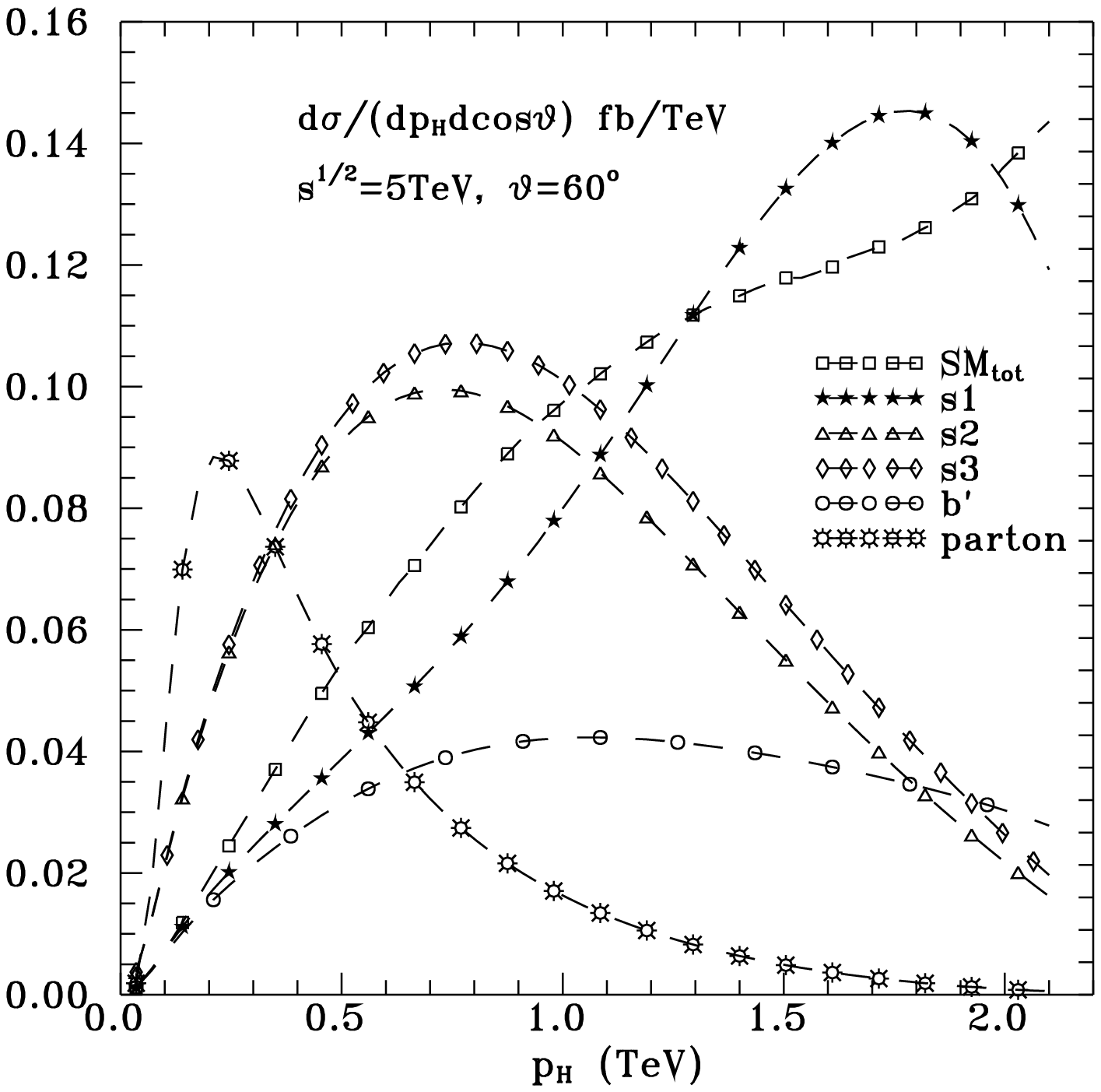,height=6.cm}
\]
\caption[1]{The  SM unpolarized differential cross sections (\ref{sigma2})
together with the new physics contributions s1, s2, s3,  $b'$, and the parton-ones
discussed in Section 5.
 The SM results in the upper panels only involve sSM, while those at the lower panels
 involve  ${\rm SM_{tot}}$; compare Figs.\ref{Gr4}.
 Again, left panels correspond to  $\sqrt{s}=1$TeV, and right panels to $\sqrt{s}=5$TeV. }
\label{Gr5}
\end{figure}

We next turn to the illustration of the above new physics effects and  compare
them to the related SM contributions. These are presented in Figs.\ref{Gr5},...\ref{Gr8}.
As already stated, with the  choice of the new physics couplings we have made for the various models,
the   differences among the various new physics models we discuss,
 are concentrated in  the shapes of the various distributions.

In Figs.\ref{Gr5} first, we present the contributions
of the scalar $s1, s2, s3$, the fermion $b'$ and the parton model discussed above,
to the unpolarized differential cross sections. In addition,
we also show in the upper and lower panels, the sSM (only the s-channel contributions)
and SMtot (both s-channel and t,u channel contributions) standard model cases
respectively.
One can immediately see the differences in the shape
and magnitude of the new contributions (s1, s2, s3, $b'$, parton) with respect to
those of the SM contribution, especially for high energy and  $p_H$.

\begin{figure}[b]
\vspace{-1cm}
\[
\epsfig{file=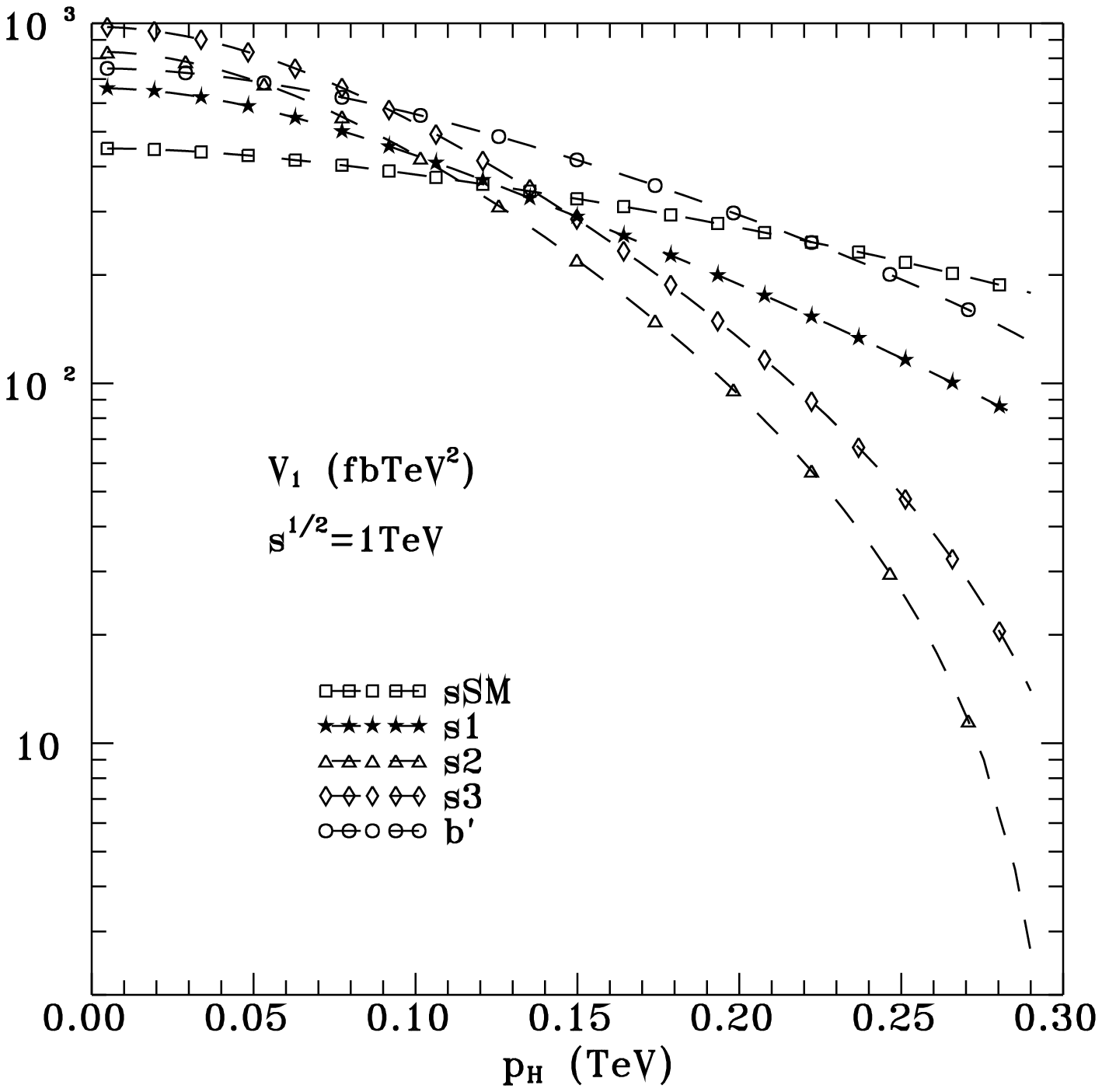, height=6.cm}\hspace{1.cm}
\epsfig{file=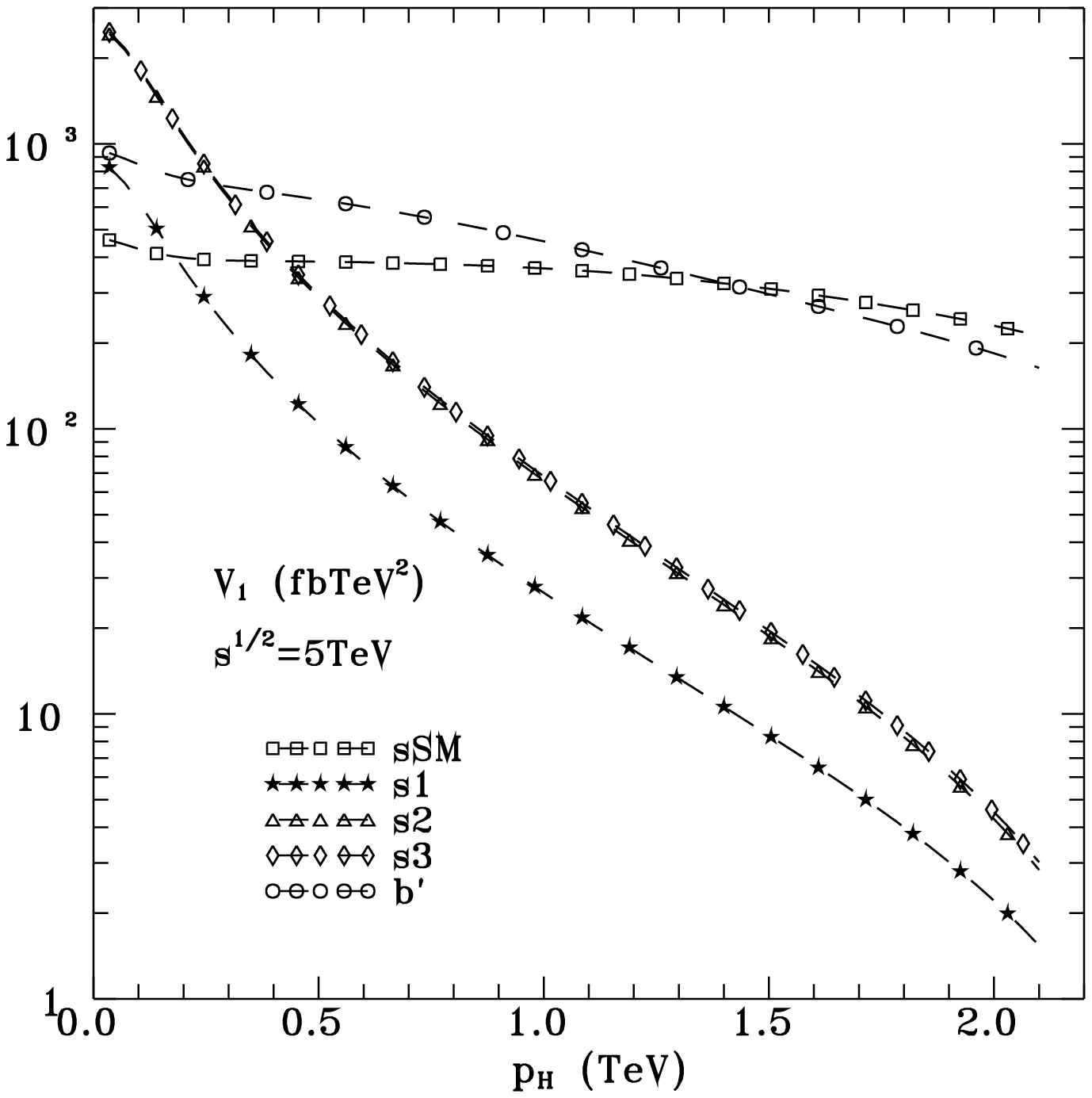,height=6.cm}
\]
\[
\epsfig{file=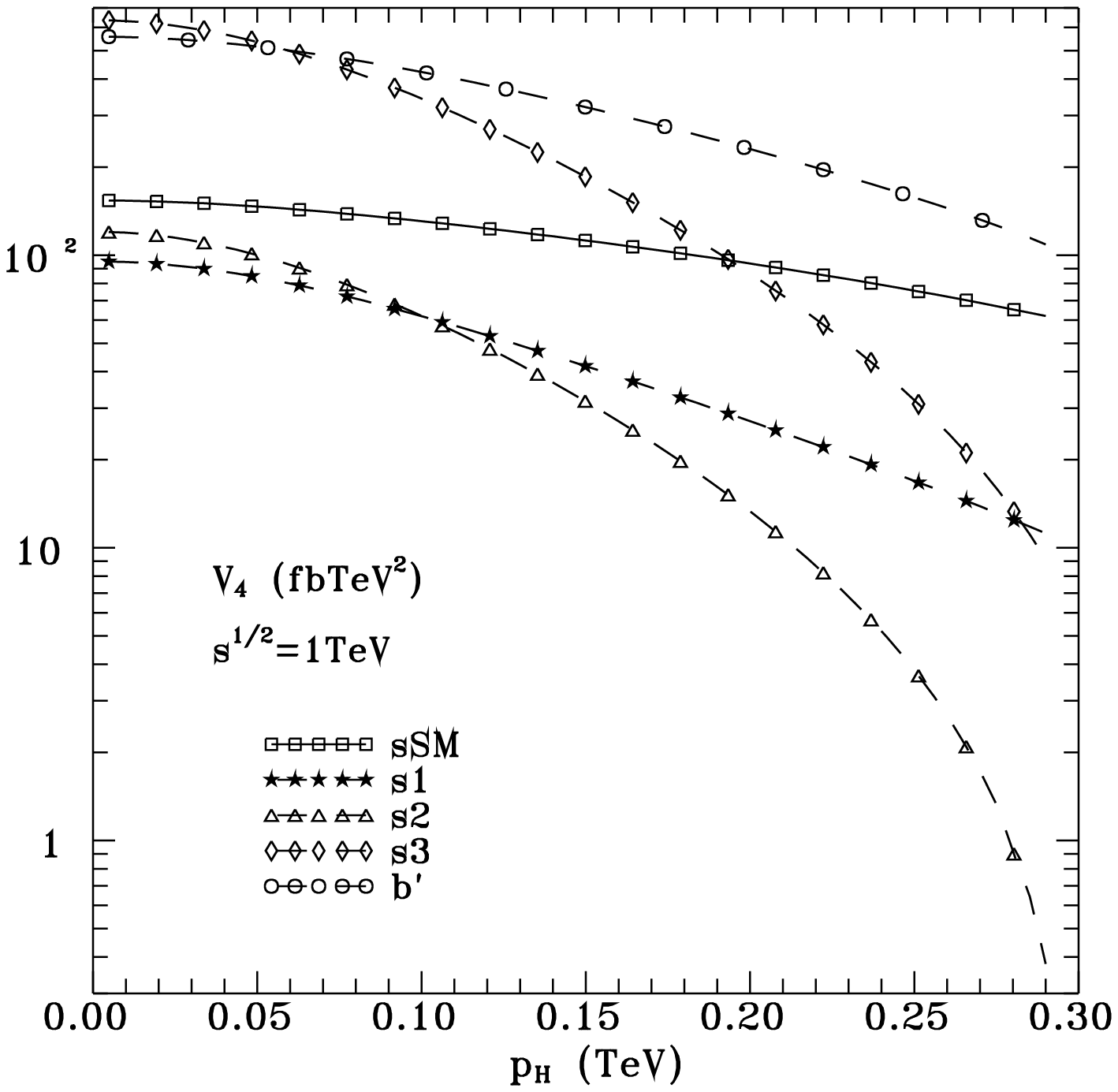, height=6.cm}\hspace{1.cm}
\epsfig{file=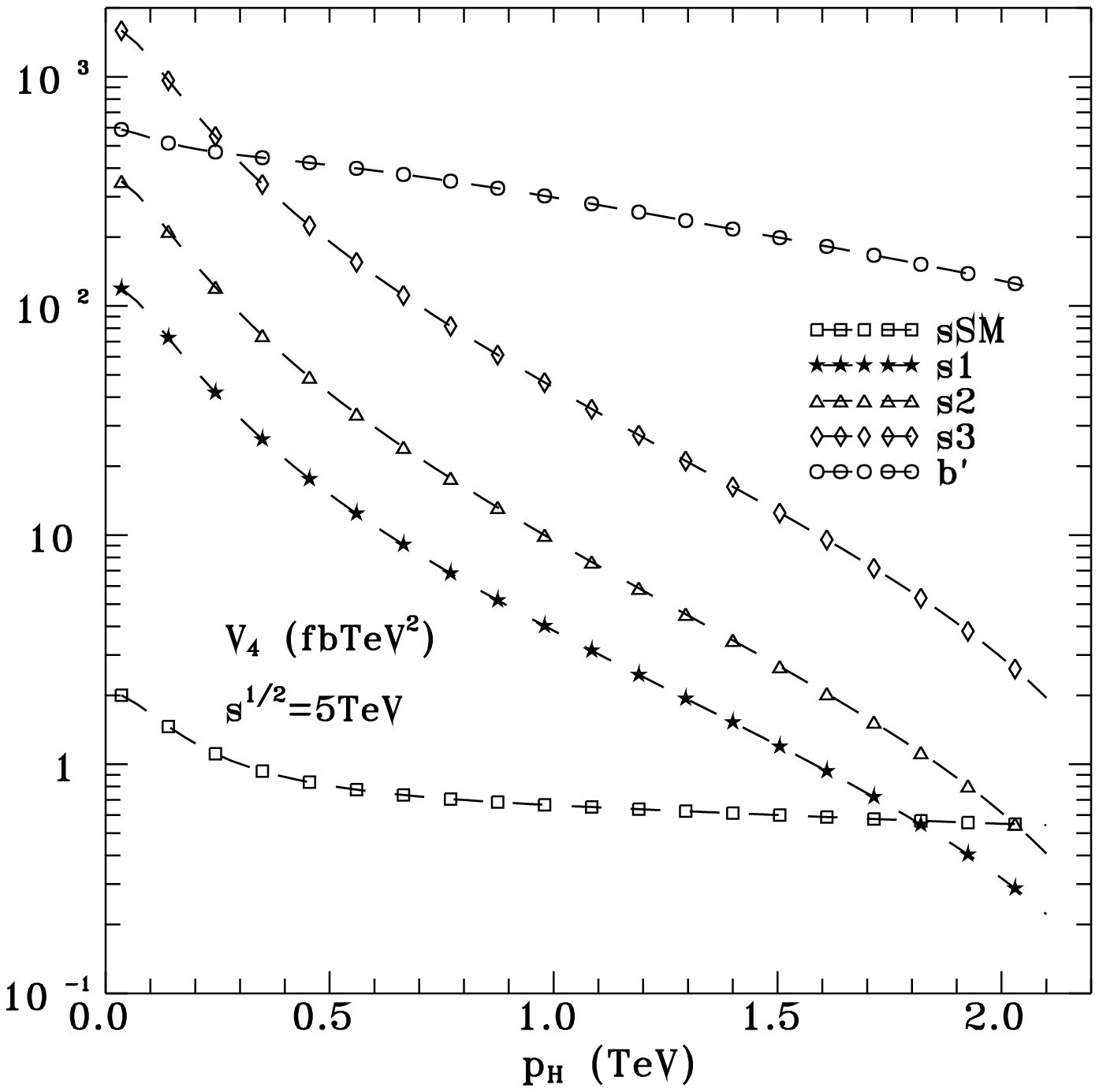,height=6.cm}
\]
\caption[1]{Structure functions $V_1$  and $V_4$ measured in ${\rm fb TeV^2}$,
for the s-channel forms sSM (see Figs.\ref{Gr4}), together with
the new physics  forms s1, s2, s3, $b'$ and the parton contributions.
Left panels correspond to  $\sqrt{s}=1$TeV, and right panels to $\sqrt{s}=5$TeV.}
\label{Gr6}
\end{figure}

\begin{figure}[t]
\vspace{-1cm}
\[
\epsfig{file=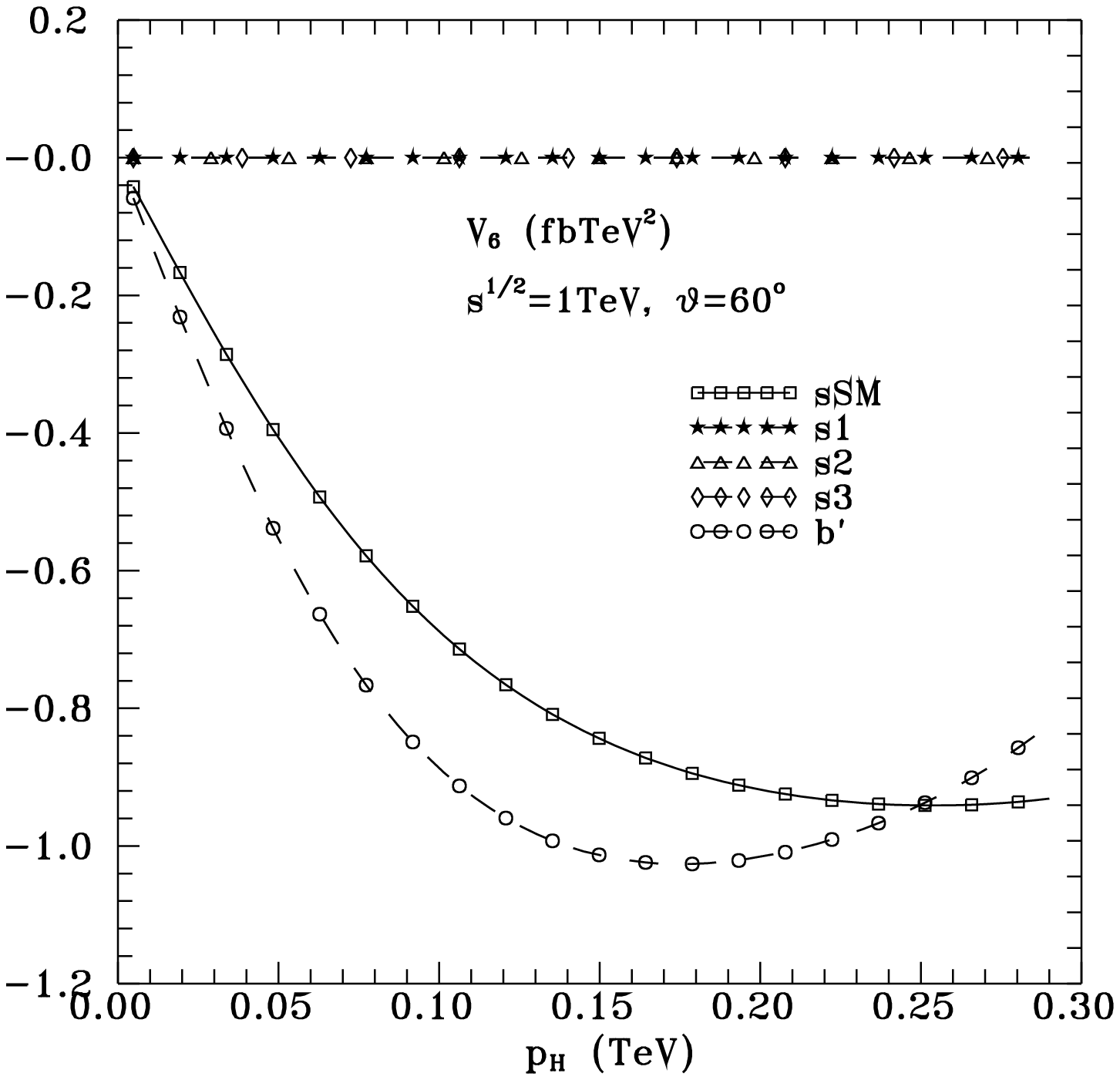, height=6.cm}\hspace{1.cm}
\epsfig{file=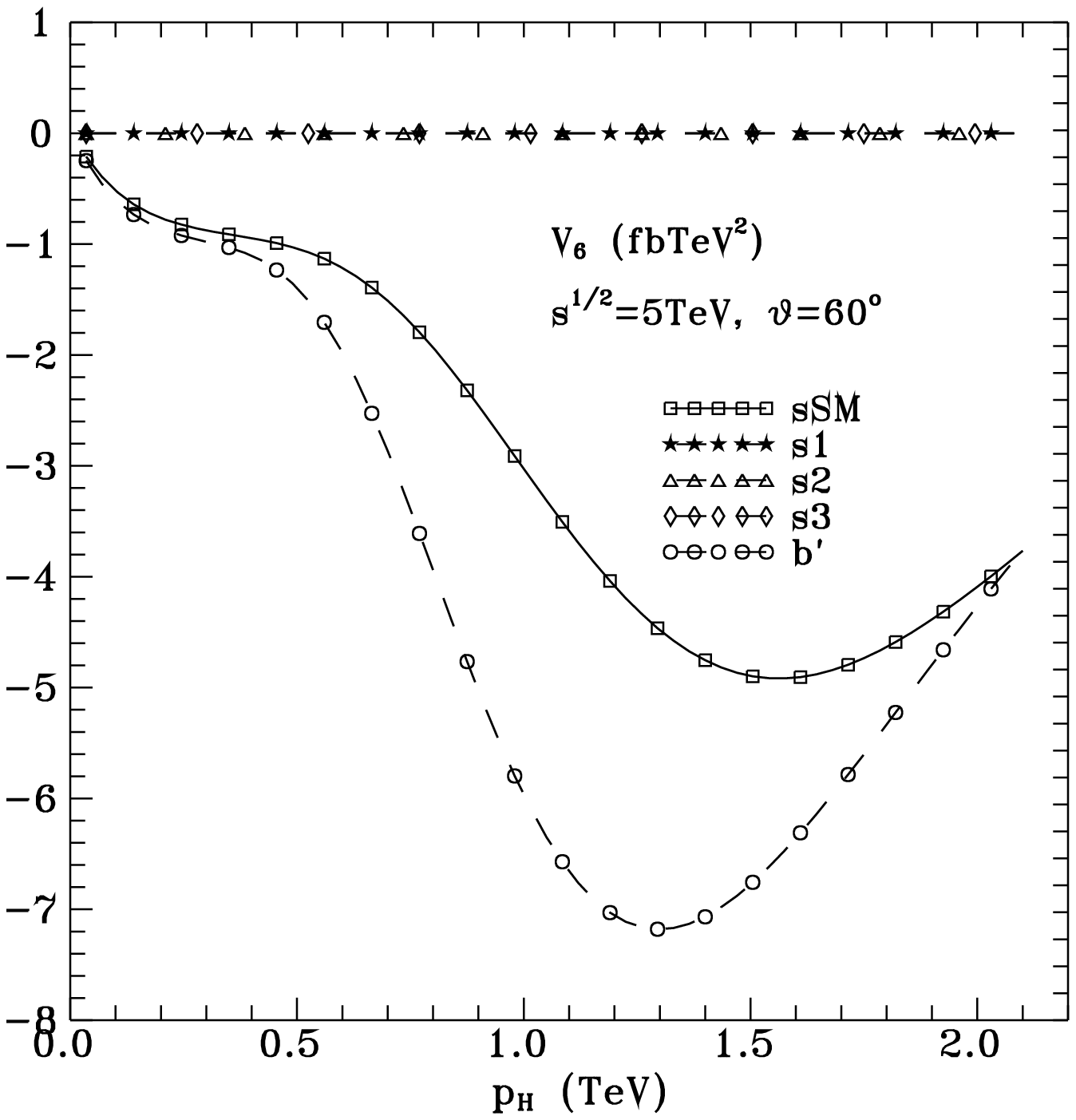,height=6.cm}
\]
\[
\epsfig{file=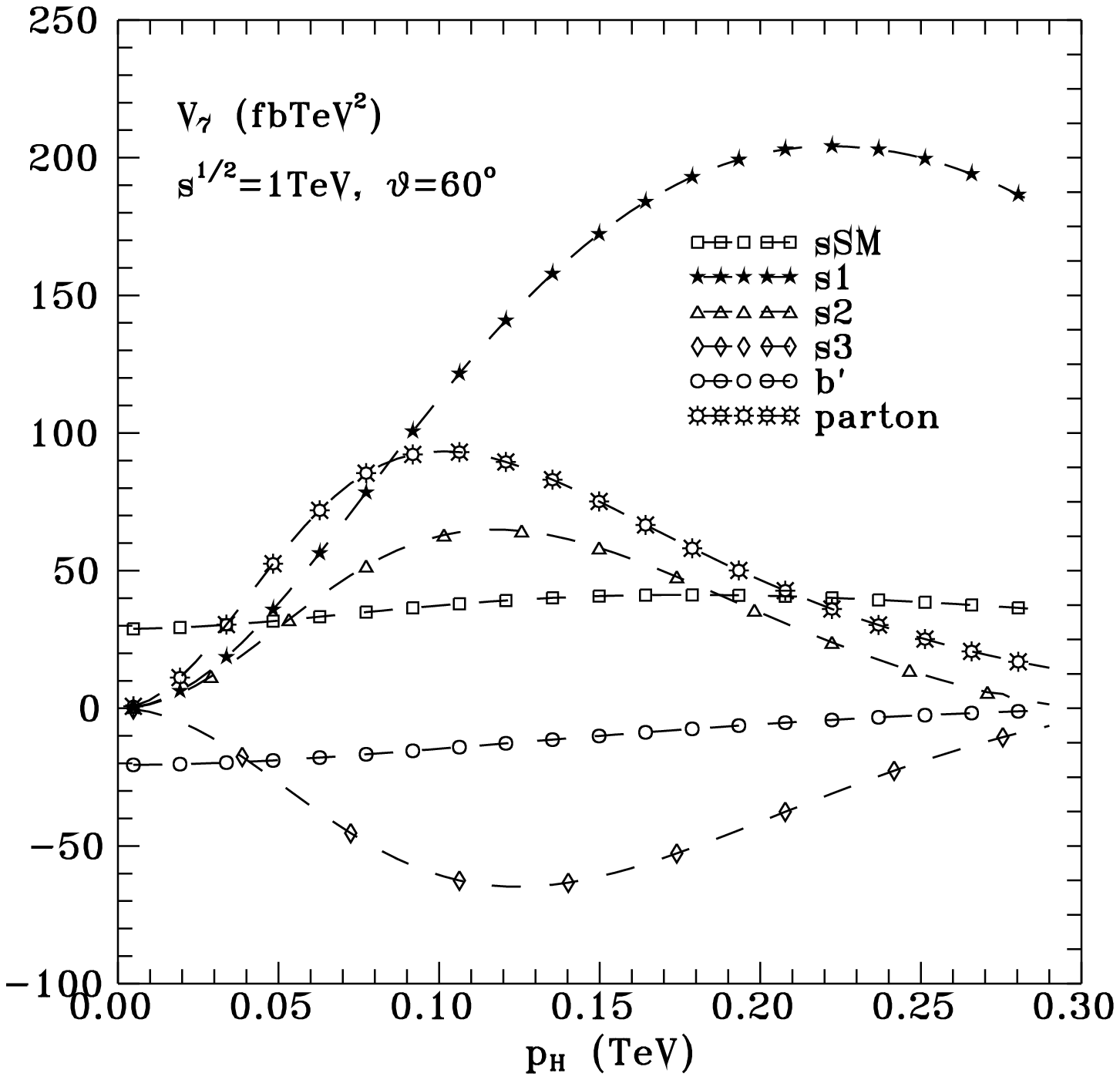, height=6.cm}\hspace{1.cm}
\epsfig{file=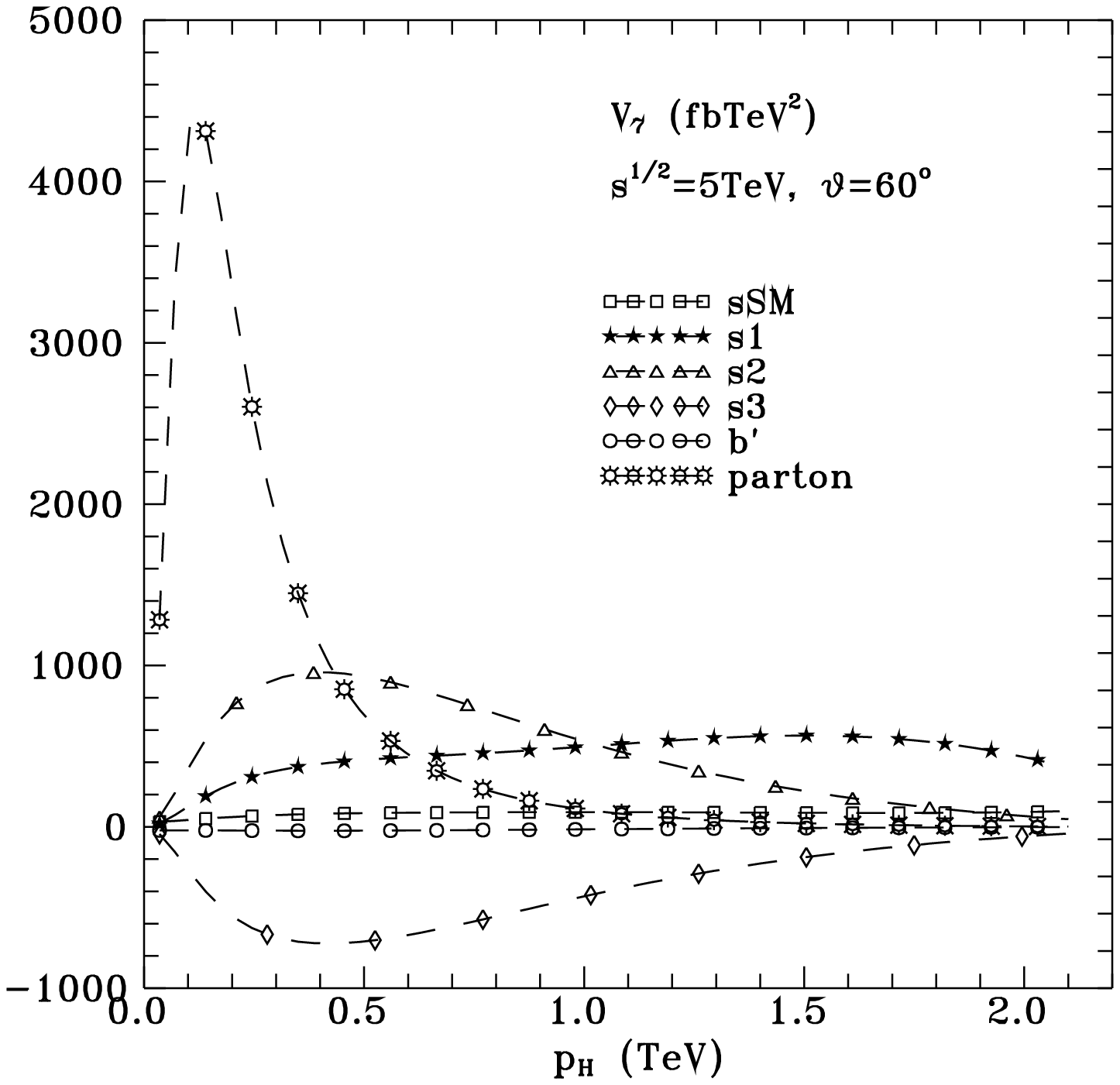,height=6.cm}
\]
\caption[1]{ As in Fig.\ref{Gr6},  for  $V_6,~V_7$,
shown respectively in the upper and lower panels.}
\label{Gr8}
\end{figure}

The origins of these differences can be analyzed by looking at the various
$V_i$ combinations in Figs.\ref{Gr6}, ...\ref{Gr8}.
The effects are much more spectacular than in the unpolarized cross section.
Each new contribution gives a specific modification of the $V_{1,...7}$.

The spin is one reason for these differences;
at high energy and $p_H$, the scalars (including the scalar parton model) do not
contribute to $V_{1,3,6}$. The masses are another reason, as one can see by
comparing the shapes of $s1$ and $s2,s3$. This feature could be useful
for identifying invisible matter.

\begin{figure}[h]
\vspace{-1cm}
\[
\epsfig{file=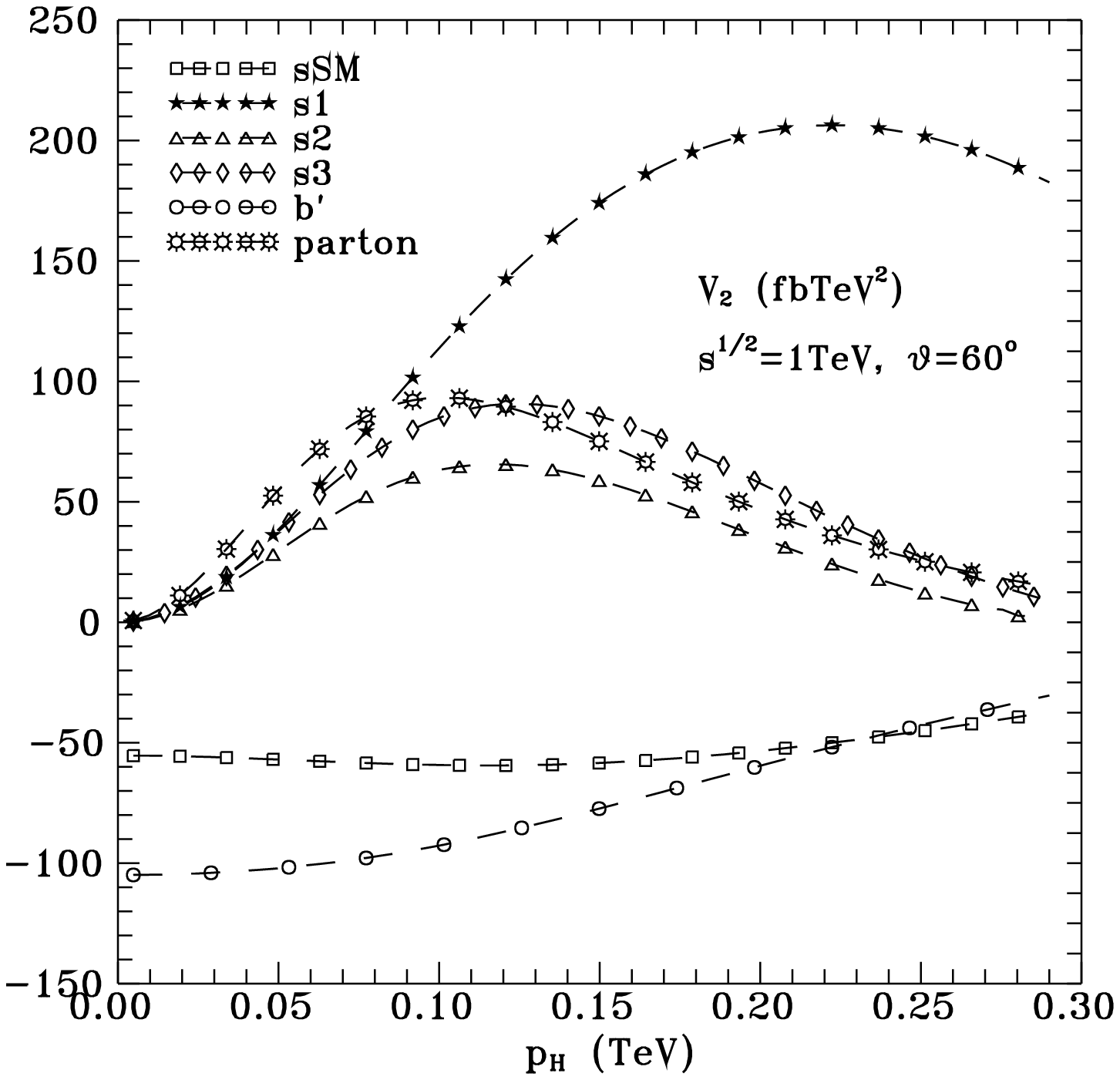, height=6.cm}\hspace{1.cm}
\epsfig{file=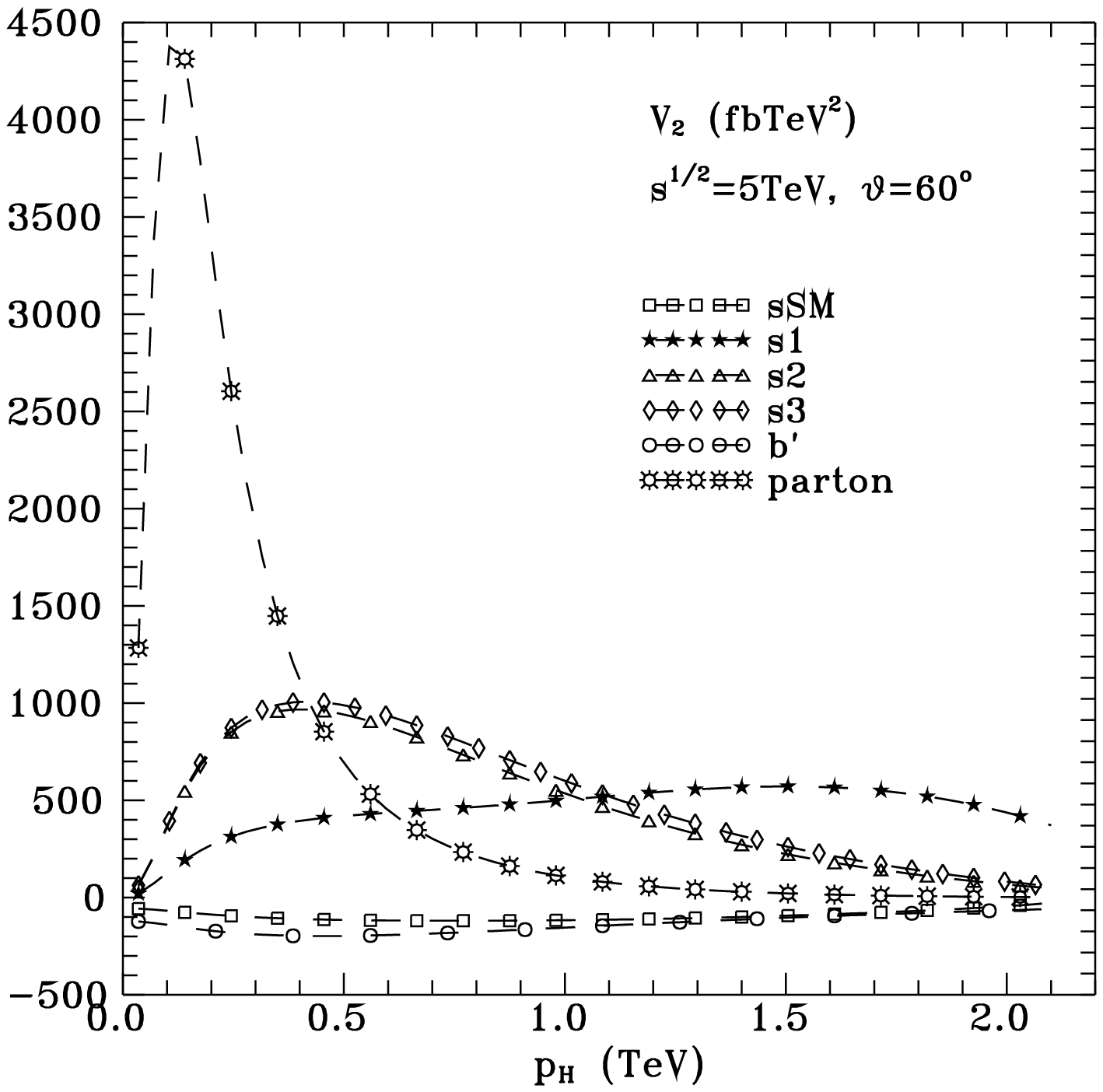,height=6.cm}
\]
\[
\epsfig{file=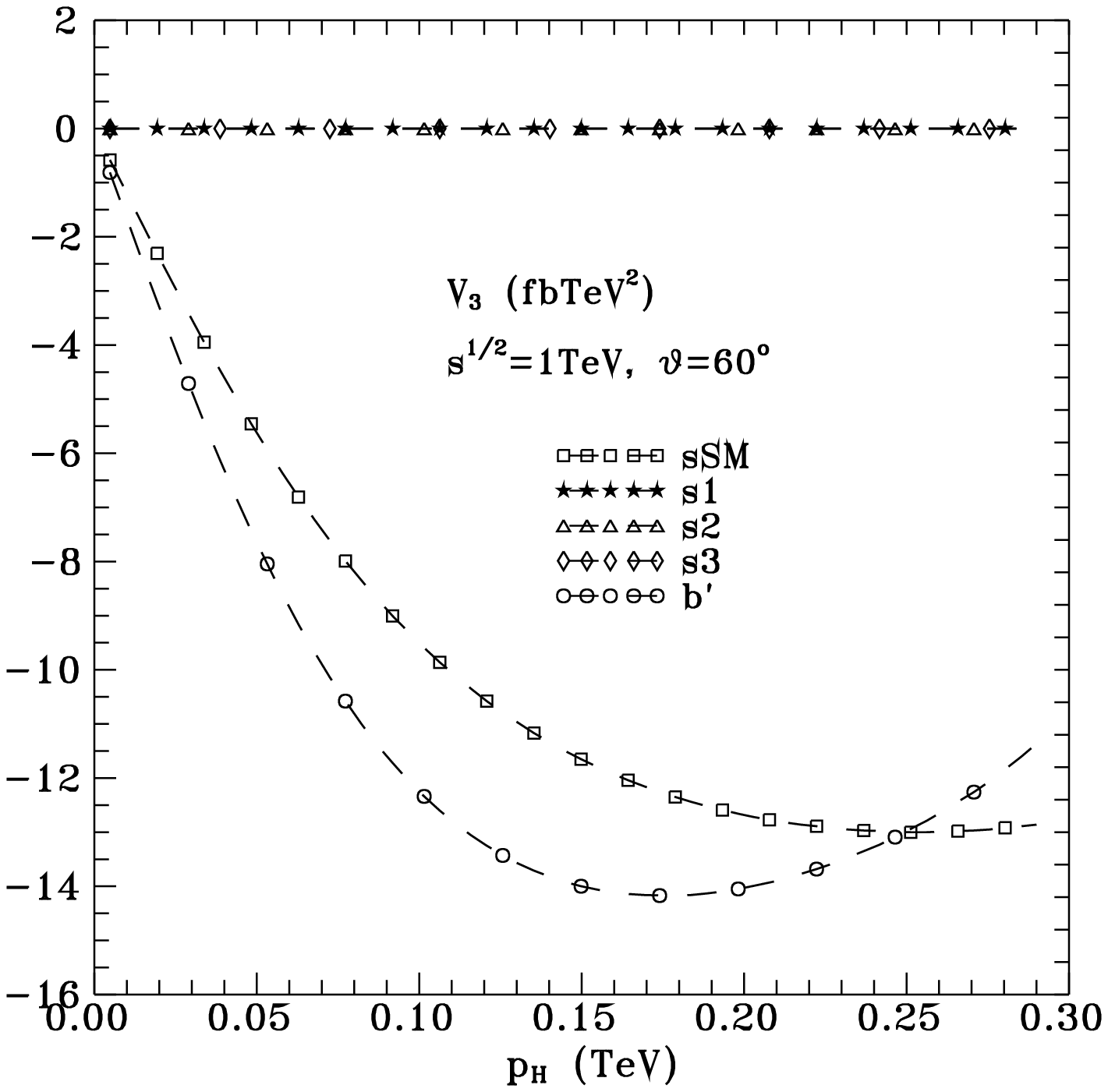, height=6.cm}\hspace{1.cm}
\epsfig{file=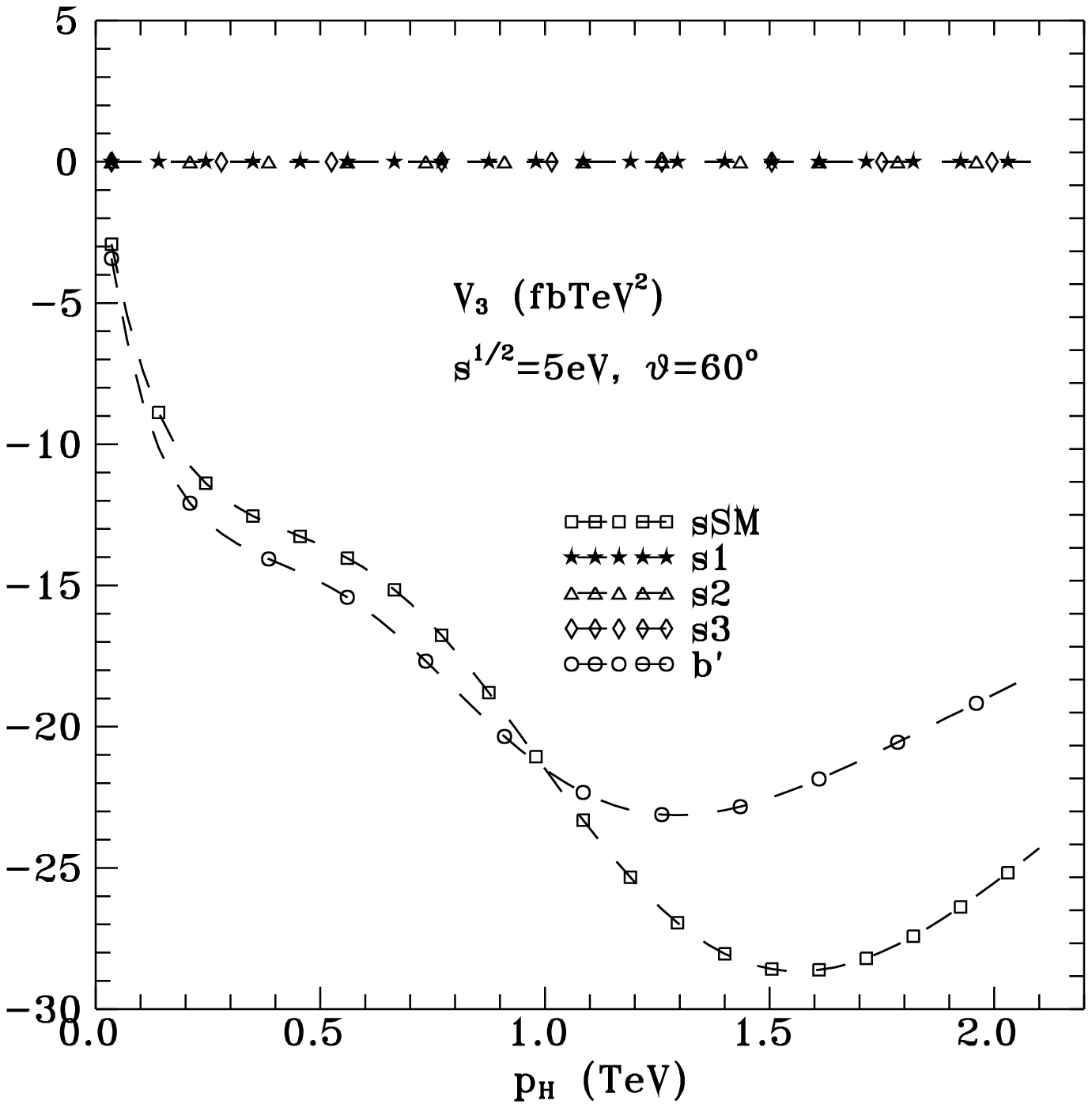,height=6.cm}
\]
\[
\epsfig{file=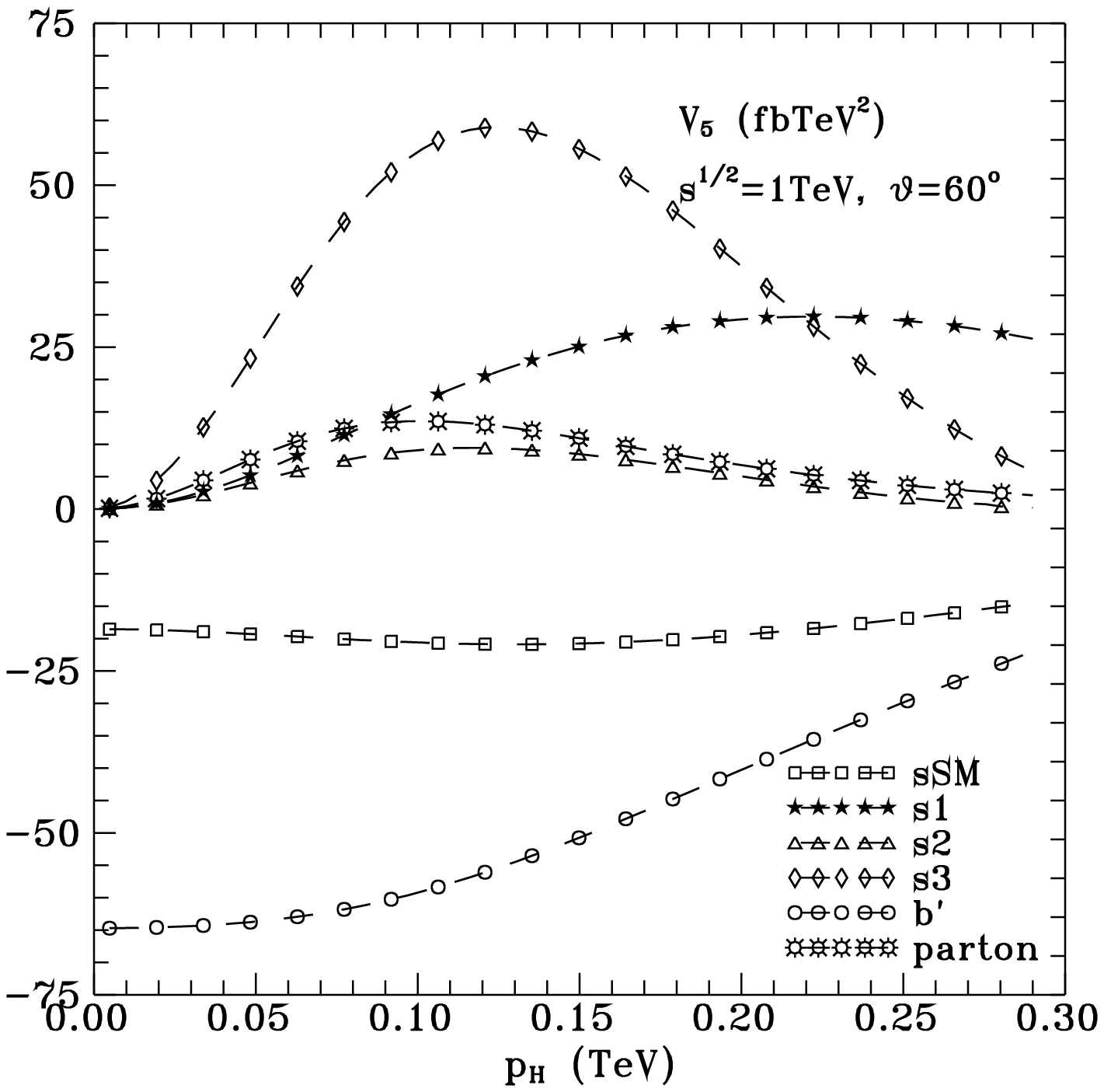, height=6.cm}\hspace{1.cm}
\epsfig{file=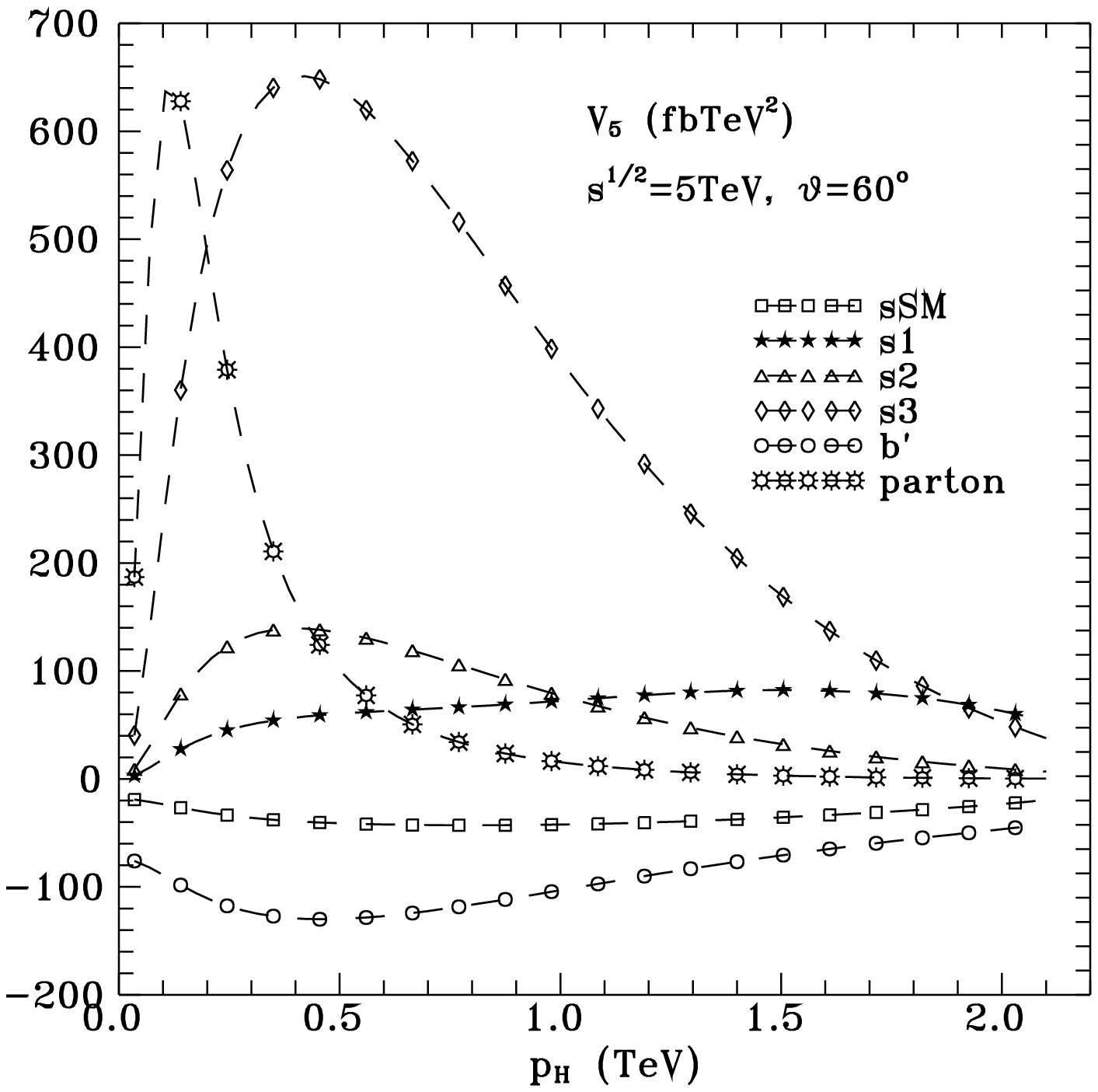,height=6.cm}
\]
\vspace{-0.5cm}
\caption[1] {As in Fig.\ref{Gr6},  for  $V_2,~V_3,~V_5$,
shown respectively in the upper, middle and lower panels.}
\label{Gr7}
\end{figure}

 In more detail, Figs.\ref{Gr6} present the angular independent structure functions $V_1,~V_4$
as functions of $p_H$. These are usually the largest and they are positive for all  $p_H$.
In contrast to them, the structure functions $V_2,~V_3,~V_5$ (Figs.\ref{Gr7}) and
$V_6,~V_7$ (Figs.\ref{Gr8}) may change sign as $p_H$ varies, and are usually considerably smaller.
Exceptions appear for the parton contributions at $\sqrt{s}=5$TeV and low $p_H$, for $V_2,~V_5,~V_7$;
and also for the  s3-contribution to $V_5$ at $p_H\sim 0.5$TeV.

\section{Conclusions and possible future developments}

In this paper we propose an analysis of the Higgs boson structure functions in
$e^-e^+$ collisions.

In doing so, we have first computed the theoretical expectations for the
SM contributions to these structure functions
and to their combinations appearing in the angular terms of the inclusive
cross section $e^-e^+ \to H + {\rm ``anything"}$.
We have then given examples of possible new particle contributions and examined how they
modify the shapes of the distributions of the various Higgs boson structure functions,
such that one can suspect what type of new particles are produced, without observing them.

These differences arise from the spins, masses and couplings of the new particles involved,
 and reflect in the  $p_H$ dependencies  and the signs and magnitudes of the various $W_i$ or
of their effective combinations $V_i$. They  are rather spectacular, first in the global
unpolarized differential cross sections
and more drastically in the various $V_i$ combinations controlling its
angular components (see illustrations). The shapes of these distributions
may  allow to guess what type of new particle (spin, mass)
is responsible of a departure from SM when it is observed, even if this particle
is not visible.

Our conclusion is then the following:
The shapes of the Higgs boson structure functions can tell us something about new physics.

The output of this first work is
essentially  a suggestion for further more detailed phenomenological and experimental
studies. Obviously several improvements
could be performed; like a complete treatment including
higher orders and background processes.

There should also be the possibility of studies of other inclusive processes.
The closest one to the case considered in the present work being for example
$q\bar q \to H+{\rm ``anything"}$, in particular
$q\bar q'\to  W\to H+{\rm "anything"}$ at LHC.
But other processes, for example with
initial photons or gluons, $\gamma\gamma \to H+{\rm ``anything"}$,
$gg \to H+{\rm ``anything"}$ could also be considered.\\

\clearpage

\clearpage

\clearpage

\clearpage

\clearpage

\clearpage

\clearpage

\end{document}